\documentclass[aps,prb,superscriptaddress]{revtex4}

\usepackage{graphics} 
\usepackage{graphicx}
\usepackage{amsmath}
\usepackage{amsfonts}
\usepackage{dcolumn}

\newcommand{\Sh}{Schr\"odinger{ }}

\newcommand{\bs}[1]{\boldsymbol{#1}}

\newcommand{\Fig}[1]{Fig. \ref{#1}}
\newcommand{\braket}[3]{\left\langle #1 \left\vert #2
            \right\vert #3 \right\rangle}
\newcommand{\brakete}[3]{\left\langle #1 \right\vert #2
            \left\vert #3 \right\rangle}
\newcommand{\brak}[2]{\left\langle #1 \left\vert
             #2 \right. \right\rangle}

\newcommand{\fues}[1]{\left(#1\right)}

\newcommand{\yav}[1]{\left[#1\right]}

\newcommand{\llal}[1]{\left\{#1\right.}
\newcommand{\rral}[1]{\left. #1\right\}}
\newcommand{\abs}[1]{\left\vert#1\right\vert}

\newcommand{\Eq}[1]{Eq. (\ref{#1})}
\newcommand{\Eqs}[1]{Eqs. (\ref{#1})}

\begin{document}  

\title[]{ Kubo formula for Floquet states and  photoconductivity oscillations in a 2D electron gas. }  
\author{Manuel Torres}
\email[Email:]{torres@fisica.unam.mx} 
\affiliation{Instituto de F\'{\i}sica,
Universidad Nacional Aut\'onoma de M\'exico, Apartado Postal
20-364,  M\'exico Distrito Federal 01000,  M\'exico}  
 
\author{Alejandro Kunold } 
\email[ Email:]{akb@correo.azc.uam.mx} 
\affiliation{Departamento de Ciencias B\'asicas, Universidad Aut\'onoma
Metropolitana-Azcapotzalco, Av. San Pablo 180,  M\'exico D. F. 02200, M\'exico}  

\date{\today}  

\begin{abstract}
The recent discovery of  the microwave  induced vanishing resistance states  in a two dimensional electron system (2DES) is  an unexpected and  surprising phenomena. In these experiments the magnetoresistance of a  high mobility  2DES under the influence of microwave radiation of frequency $\omega$  at moderate values of the magnetic field, exhibits strong oscillations  with  zero-resistance states (ZRS)   governed by the ratio $\omega /\omega_c$, where $\omega_c$ is the cyclotron frequency. 
In this work we present a  model for the photoconductivity of a  two dimensional electron system (2DES) subjected  to a magnetic field. The model includes the microwave and Landau contributions in a non-perturbative  exact way, impurity scattering   effects are treated perturbatively.  In our model,  the Landau-Floquet  states act coherently with respect to the  oscillating field of  the impurities, that  in turn induces  transitions between these levels.  Based on this formalism, we provide a  Kubo-like   formula that takes into account the oscillatory Floquet structure of the problem. We study the effects of  both short-range and long-range disorder on the photoconductivity.  Our calculation yields a magnetoresistance oscillatory behavior with  the correct period and phase. It is found that, in agreement with experiment,  negative dissipation can only be induced in very high mobility samples.  We analyze the  dependence of the results  on the microwave  power and polarization. For high-intensity radiation multi-photon processes  take place predicting new negative-resistance states centered  at  $ \omega / \omega_c=1/2$, and  $ \omega / \omega_c= 3/2$. 

\end{abstract}
\pacs{73.43.-f,72.40.+w,75.47.-m}
\maketitle

\section{Introduction.}\label{intro}

Two-dimensional  electron systems $(2DES)$  in a perpendicular  strong magnetic field have been extensively  studied  in relation with the quantum Hall effect.  Recently, two experimental groups \cite{zudov1,mani1,zudov2,mani2},  reported the observation of a novel phenomenon: the existence  of  zero-resistance states (ZRS)   in an ultraclean  $GaAs/Al_xGa_{1-x} As$ sample  subjected to microwave radiation and moderate magnetic fields.      The   magnetoresistance  exhibits strong oscillations  with ZRS  governed by the ratio $\epsilon=  \omega/\omega_c$, where $\omega_c$ is the cyclotron frequency.  According  to  Zudov $et \, al.$,  the   oscillation amplitudes  reach  maxima  at $\epsilon= \omega/\omega_c =j$ and minima at  $\epsilon =j + 1/2$, for $j$ an integer. On the other hand Mani $et \, al.$  reported also  a periodic oscillatory behavior, but with maxima at $\epsilon =j - 1/4 $ and minima at $\omega/\omega_c =j +1/4$.  Additionally experimental work appeared  recently \cite{doro,willett,zudov3,kova,stud}. 
 
In spite of a large number of theoretical works \cite{ry1,ry2,durst,andre,ander,shi,lei,vavilov,dmi1,dmi2,kennett},  a complete understanding of the new  effects in 2DES induced by microwaves  has not yet been achieved.
A pioneering work put forward by  Ryzhii \cite{ry1,ry2} predicted the existence of negative-resistance 
  states (NRS). Durst and collaborators \cite{durst}  found also NRS  in a  a diagrammatic calculation of the photoexcited electron scattered by a disorder  potential.  A possible connection between the calculated NRS  and the observed vanishing resistance was put forward in reference \cite{andre}, noting that a general analysis of Maxwell equations shows that negative resistance  induces an instability  that  drives  the system  into a  ZRS.  Whereas some of the models \cite{ry1,ry2,durst,vavilov} are  based on an impurity assisted mechanisms,  there are alternative explanations \cite{doro,dmi2,kennett}  in which the leading contribution arises from the modifications of the electron distribution function induced by  the microwave radiation. These models, as well as some of their predictions remain to be tested experimentally.

  In this work we present a  model which includes the Landau and radiation contribution  (in the long-wavelength limit) in a 
  non-perturbative  exact way.  Impurity scattering   effects are treated perturbatively. With respect to the  Landau-Floquet states, the impurities act as  a coherent oscillating field which  induces the  transitions that  proved to be essential in order to reproduce the observed  oscillatory behavior of  the magnetoresistance. Based on this formalism a  Kubo-like  expression for the conductance  is provided.  Our results    display  a  strong oscillatory behavior  for $\rho_{xx}$  with NRS.    It is found  that $\rho_{xx}$ vanishes at $\epsilon=   \omega /\omega_c =j$ for $j$ integer.  The oscillations follow a pattern with minima at $\epsilon =j + \delta $, and maxima at   $\epsilon =j -  \delta  $, adjusted with $\delta \approx 1/5$.  The model is  used to test chirality  effects induced by the magnetic field, calculations are carried out  for various $\bs E$-field polarization's.  Finally,  we explore the nonlinear regime in which multi-photon processes   play an essential role. 
   
 The paper is organized as follows. In the next section we present
the model and the   method that allow us to obtain the exact solution of the Landau-microwave
system, as well as  the perturbative corrections induced by the impurity potential. 
 In section \ref{kubo} we develop the formulation of $dc$ electrical linear response theory  valid in arbitrary  magnetic field and microwave radiation.    A discussion of relevant numerical calculations is presented in section (\ref{results}). The last section contains a summary of our main results.  
Details of the calculations are summarized in the appendices.

\section{The Model.}\label{model}
We consider the motion of an  electron in two dimensions subject to a uniform magnetic  field  $\mathbf{B}$  
perpendicular to the plane  and driven by  microwave radiation. In the long-wave limit the 
 dynamics is governed by the \Sh equation 
\begin{equation}\label{ecs1}
i \hbar \frac{\partial \Psi }{\partial t}= H \Psi  =  \left[  H_{\{B,\omega\}}  + {\tilde V}({\bs r} )  \right] \Psi  \, , 
\end{equation}
here $H_{\{B,\omega\}}$ is the Landau hamiltonian coupled to the radiation  field (with $ \lambda \to \infty $) and ${\tilde V}({\bs r})$ is any  potential that can can be decomposed in a Fourier expansion.  The method applies in general if    ${\tilde V} ({\bs r})$  includes various possible effects such as: lattice periodic potential, finite wave-length corrections, impurity scattering, etc; however as it will  be lately argued, the impurity scattering is the most likely explanation for the recent experimental results.  One important remark with relation to the impurity  potential in \Eq{ecs1} is that to start with,  it  should only include the polarization effects produced by the combined effects of the Landau-Floquet states and the impurity potentials, the broadening effects produced by this potential  are, as usually, included through  the Kubo formula. Then we write 
${\tilde V}(\bs r)$ as
\begin{equation}\label{potren}
 {\tilde V}(\bs r) =  V(\bs r)  - \Delta  V(\bs r), \hskip2.0cm   \Delta  V(\bs r)= W^{\dag} \,  V (\bs r) W \, , 
\end{equation}
were  $W$ is the transformation  that   takes  exactly into account the 
microwave-Landau dynamics, it is  explicitly given in \Eq{opw}. In the absence of the  microwave radiation, 
$W \equiv 1$ and  ${\tilde V}(\bs r) $ vanishes.  The impurity scattering potential $V(\bs r) $ is decomposed as

\begin{equation}\label{potimp}
 V(\bs r) = \sum_i \int \frac{d^2 q}{(2 \pi)^2} \, V( \bs q) \exp\{i {\bs q}  \cdot \left( {\bs r }- {\bs r}_i \right)   \} \, , 
\end{equation}
here $ {\bs r}_i$ is the position of the $i$th impurity and  the explicit form of $V( \bs q) $  depends of the mechanism that applies under  particular physical 
conditions, some examples will be consider in section  (\ref{results}). 
$H_{\{B,\omega\}}$ is then written as 
\begin{equation}\label{ham1}
H_{\{B,\omega\}} = \frac{1}{2m^*} {\bs \Pi}^2 \, ,  \hskip3.5cm   {\bs \Pi} = \mathbf{p}+e\mathbf{A} \, , 
\end{equation}
$m^*$ is the effective electron mass and  the vector potential  $\mathbf{A}$  includes  the external magnetic  and  radiation fields (in the $\lambda \to \infty $ limit)
 contributions 
  \begin{equation}\label{vecpot}
\mathbf{A} = - \frac{1}{2} \bs r \times \bs B  +  Re \,\, \left[  \frac{\bs {E}_0}{\omega} \exp\{ -i \omega t \} \right] \, . 
\end{equation}
   We first consider the exact solution of the microwave driven Landau problem, the impurity scattering effects are  lately 
 added perturbatively. This approximation is justified on  the following conditions:  $(i)$:   $ \vert  V \vert /  \hbar \omega_c << 1$ and   $(ii)$: $\omega \, \tau_{tr}   \sim  \omega_c \, \tau_{tr}  >>  1$; $\tau_{tr}$  is the transport  relaxation time that is estimated using its relation to the 
 electron mobility $\mu = e \tau_{tr} / m^*$.  As discussed in section  (\ref{results})   both conditions are  fully complied.

 The system  posed by $H_{\{B,\omega\}}$  can be recast as a forced  harmonic oscillator,  a 
 problem that was solved long time ago by Husimi \cite{husi}.  Following the formalism developed in references \cite{Kunold1,Torres1}, we introduce a canonical transformation to new variables 
$Q_\mu,P_\mu$; $\mu=0,1,2$,  
  according to 
\begin{align}\label{trancan}
Q_0&=  t     \, , &P_0 = i \partial_t + e \phi + e {\bs r} \cdot  {\bs E}  ,\nonumber \\
\sqrt{e B} Q_1&=\Pi_y   \, , & \sqrt{e B}P_1 =\Pi_x ,\nonumber \\
\sqrt{e B} Q_2&=p_x+ e A_x +  e By  \, , &\sqrt{e B} P_2=p_y+ e A_y - e B x.
\end{align}
It is easily verified that the transformation is indeed  canonical, 
the  new variables obey  the commutation rules:  $- \left[Q_0,P_0\right]=  \left[Q_1,P_1\right]= \left[Q_2,P_2\right]=i $; all other commutators being zero.
The inverse transformation gives
 \begin{equation}\label{invtran}
x=  l_B  \left(Q_1-P_2 \right)  , \hskip2.0cm  y=  l_B   \left( Q_2-P_1 \right) ,
\end{equation}
where $l_B=  \sqrt{\frac{\hbar}{eB}}$ is     
the magnetic length. 
 The operators $( Q_2 , P_2)$ can  be  identified with the generators of the electric-magnetic translation symmetries \cite{Ashby1,Kunold2}.
Final results are independent  of the selected gauge. 
From the operators in \Eq{trancan}   we construct two pairs of
harmonic oscillator-like ladder operators: $(a_1,a_1^{\dag})$, and $(a_2,a_2^{\dag})$ with:
\begin{equation}\label{opasc}
a_1 =   \frac{1}{\sqrt{2}}    \left(P_1-iQ_1\right), \hskip2.0cm    a_2 =  \frac{1}{\sqrt{2} }    \left(P_2 -i Q_2\right), 
\end{equation}
obeying:  $[a_1,a_1^{\dag}]=[a_2,a_2^{\dag}]=1$, and $[a_1,a_2]=[a_1,a_2^{\dag}]=0$.

It  is now  possible to find a unitary transformation that exactly diagonalizates    $H_{\{B,\omega\}}$, it yields
\begin{equation}\label{tran1}
W^{\dagger} H_{\{B,\omega\}}  W = \omega_c \left( \frac{1}{2} + a_1^\dagger \, a_1\right) \equiv H_0 \, , 
\end{equation}
with the cyclotron frequency $\omega_c = eB/m^*$ and the $W(t)$ operator     given by 
\begin{equation}\label{opw}
W(t)=  \exp\{i \eta_1 Q_1\}  \exp\{i \xi_1 P_1\} \exp\{i \eta_2 Q_2\}  \exp\{i \xi_2 P_2\}   \exp\{- i  \int^t  {\mathcal L} dt^\prime \} 
\, , 
\end{equation}
where  the functions $\eta_i(t)$ and $\xi_i(t)$ represent the solutions to the classical equations of motion that follow from the variation of the Lagrangian 

\begin{equation}\label{lagrang}
{\mathcal L}  =  \frac{\omega_c}{2} \left( \eta_1^2 + \zeta_1^2 \right) +    \dot \zeta_1\eta_1 +  \dot \zeta_2 \eta_2 
+  e l_B \,    \left[ E_x  \left( \zeta_1 + \eta_2 \right)   +    E_y   \left(\eta_1 + \zeta_2  \right) \right]  
\, .
\end{equation}
The explicit  form of the solutions for   $\eta_i(t)$ and $\xi_i(t)$  are given in the appendix (\ref{apenA}). 

Let us now  now consider the complete Hamiltonian   including  the  contribution from the   ${\tilde V}(\bs r)$  potential. When  the transformation induced by  $W(t)$ is applied  the   Schr\"odinger equation in (\ref{ecs1})  becomes
\begin{equation}\label{ecs2}
P_0 \Psi^{(W)}  =   H_0 \Psi^{(W)} +   V_W (t) \Psi^{(W)} \, , 
\end{equation}
where $\Psi^{(W)}= W(t)  \Psi$ and 
\begin{equation}\label{pottrans}
   V_W  (t)  = W(t) {\tilde V} ( {\bs r  }) W^{-1}(t)  =  W(t)  V  ( {\bs r  }) W^{-1}(t) - V  ( {\bs r  }) \, .
\end{equation}
  Notice that the impurity potential acquires a time dependence  brought by   the  $W(t)$ transformation. The problem is now solved in the interaction representation  using first order time dependent perturbation theory. 
In the interaction representation $\Psi^{(W)}_I = \exp\{i H_0 t\} \Psi^{(W)}$, and the  Schr\"odinger equation becomes 
\begin{equation}\label{ecs3}
i \partial_t  \Psi^{(W)}_I  =  \left \{ V_W (t) \right \}_I  \Psi^{(W)}_I  \, . 
\end{equation}
The equation is solved in terms of the evolution operator $U(t)$,  in such a way  that $\Psi^{(W)}_I  (t) = U(t - t_0) \Psi^{(W)}_I  (t_0)$.
The solution of the evolution operator in first order perturbation theory is given by the expression 
\begin{equation}\label{opu}
U(t)  = 1 - i \int_{-\infty}^{t}  dt^\prime \left[ W^{\dagger}(t^\prime) {\tilde V} (\bs r) W(t^\prime)   \right]_I  \, , 
\end{equation}
that  is explicitly evaluated in the appendix (\ref{apenB}). The interaction is adiabatically  turned off as $t_0 \to  - \infty$,
in which case the asymptotic  state is selected as  one of the Landau-Floquet  eigenvalues of $H_0$, $i.e.$ 
$\vert  \Psi^{(W)}_I (t_0) \rangle \to  \vert  \mu,k \rangle$. The solution to the original \Sh equation in \Eq{ecs1} has been achieved 
by means os three successive  transformations, which expressions have been explicitly obtained:
\begin{equation}\label{3trans}
\vert  \Psi_{\mu,k} (t) \rangle  = W^\dag  \,  \exp\{-i H_0 t\} \,  U(t - t_0) \,  \vert  \mu ,k \rangle .
\end{equation}
 As discussed on the appendix (\ref{apenA}) the  index $k$ labels the degeneracy of the Landau-Floquet states. Selecting the $P$-representation the dependence of the wave function on $k$ becomes  very simple, see \Eq{wf2},
for simplicity in what follows the index $k$ will not be shown. 
The expression  of the Kubo formula that will be derived in section (\ref{kubo})    requires the knowledge of   the
matrix elements of the momentum operator ${\bs \Pi}$:
\begin{equation}\label{mopmom}
 {\brakete{\Psi_\mu}{{\bs \Pi}_i}{\Psi_\nu}} = {\brakete{\mu}{U^\dag(t - t_0 ) \left[ W {\bs \Pi}_i W ^\dag  \right]_I U(t - t_0)}{ \nu}} \, .
\end{equation}
Let us first consider the term inside the square brackets,  using the explicit form of the operators in
 \Eqs{ham1}  (\ref{trancan}), and (\ref{opw})
it yields
\begin{equation}\label{tranv}
W  {\bs \Pi}_i W^\dag =
\llal{\begin{array}{ll}   \sqrt{e B} ( P_1 - \eta_1),&   \hskip1.5cm  i = x, \\
 \sqrt{e B} ( Q_1 - \xi_1) ,
& \hskip1.5cm  i = y .
\end{array}}
\end{equation}
If we now utilize the result  for the evolution operator $U$ given in  the appendix (\ref{apenB}),  we can explicitly 
work out the   matrix elements of the momentum operator
 \begin{eqnarray}\label{emopm}  
  {\brakete{\Psi_\mu}{{\bs \Pi}_i}{\Psi_\nu}} =    \sqrt{ \frac{e B}{2} } &&  \left( a_j \sqrt{\mu} e^{i \omega_c t} \delta_{\mu,\nu +1} 
+   b_j \sqrt{\nu } e^{-i \omega_c t} \delta_{\mu,\nu - 1} \right)  \nonumber  \\ 
 & & \hskip3.0cm + \, \sqrt{e B} \,  \sum_l \, \,  e^{i ( {\cal E} _{\mu \nu} + \omega l - i \eta  t)} \, \, \Delta ^{(l)}_{\mu \nu} (j) .
 \end{eqnarray} 
Here the following definitions were introduced:  $ {\cal E} _{\mu \nu} =  {\cal E} _\mu - {\cal E} _\nu$,  $a_j = b_j= 1 $ if $j=x$ and $a_j = -b_j= -i $  if $j=y$,  and   $ \Delta ^{(l)}_{\mu \nu} (j)$ is given by 
\begin{equation} \label{defdelta}
  \Delta ^{(l)}_{\mu \nu} (j) = \delta_{\mu\nu} \left[\rho_j \delta_{l,1} +  \rho^*_j \delta_{l,-1}  \right] 
  -  \frac{1}{\sqrt{2}} \left[  \frac{  a_j \tilde{q}^* C_{\mu\nu}^{(l)}}{{\cal E} _{\mu\nu} - \omega_c + \omega l - i \eta }
 +   \frac{ b_j \tilde{q} C_{\mu\nu}^{(l)}}{{\cal E} _{\mu\nu} + \omega_c + \omega l - i \eta }\right], 
\end{equation}
where $\tilde{q} = i l_B (q_x - i q_y)/ \sqrt{2}$,   and the expressions for the functions $\rho_i$,  and $ C_{\mu,\nu}^l$ are  worked out as 
\begin{equation} \label{defrho}
  \rho_1= \frac{e l_B E_0 \left(- i \omega \epsilon_x + \omega_c \epsilon_y \right)}{\omega^2 -\omega_c^2 + i \omega \Gamma_{rad}}
  \, , \hskip2.5cm  
   \rho_2= \frac{e l_B E_0 \left(\omega_c \epsilon_x + i  \omega \epsilon_y \right)}{\omega^2 -\omega_c^2 + i \omega \Gamma_{rad}}
   \, , 
\end{equation}
and
\begin{equation} \label{defc}
 C_{\mu,\nu}^{(l)}   =\sum_i  \int \frac{d^2q}{(2 \pi)^2}   V({\bs q }) e^{-i { \bs q} \cdot { \bs r}_i}D_{\mu\nu}(q) \left(\frac{\Delta}{i \vert  \Delta \vert} \right)^l
 {\tilde J}_l  \left( \vert  \Delta \vert \right) \, , 
\end{equation}
where  ${\tilde J} _l = J_l  - \delta_{l,0} $,    $J_l $  being   the   Legendre polynomials,    and  $D_{\mu\nu}(q) $ is given in terms of  the  generalized  Laguerre polynomials in (\ref{laguerres}), and 
\begin{equation} \label{defdelta2}
\Delta =  \frac{\omega_c l_B^2 e E_0 }{\omega \left( \omega^2 - \omega_c^2 + i \omega \Gamma_{rad} \right)}
\left[ \omega \left(q_x \epsilon_x + q_y \epsilon_y  \right)  + i \omega_c \left(q_x \epsilon_y  -  q_y \epsilon_x  \right) \right] \, . 
\end{equation}
  It is important to notice that the subtracted term ${\tilde J} _l = J_l  - \delta_{l,0} $, has its origin in the fact
  that the impurity potential \Eq{potren} includes only the dynamical effects, with the corresponding zero 
  field-term conveniently subtracted. This procedure is justified because the broadening effects 
  produced by $V (\bs r )$ are separately included via the Kubo formula, see appendix  (\ref{apenE}). The subtraction 
  ${\tilde J} _l = J_l  - \delta_{l,0} $  becomes  essential, otherwise the longitudinal resistance 
  would be dominated by the $l=0$ term, producing incorrect  results.

\section{Kubo formula for Floquet states.}\label{kubo}

In this section we shall develop the Kubo  formula that applies   when the  dynamics includes  Landau-Floquet states as those in \Eq{3trans}.  We take the perturbing electric field to have the form 
${\bs E}_{ext} = {\bs E}_0 \, cos \left( \Omega t\right) \, exp \left( - \eta \vert t \vert  \right)$. 
The static limit is obtained with $\Omega \to 0$ and $\eta $ represents the rate at which the perturbation is turned on and off.
The perturbing  electric field  is included in the vector potential, as we are interested in the linear response the perturbing 
potential has the form 
\begin{equation}\label{potext} 
V_{ext} = \frac{1}{m} {\bs \Pi} \cdot {\bs A}_{ext} \, , \hskip2.5cm {\bs A}_{ext}= \frac{{\bs E}_0}{\omega} \, sin \left( \Omega t\right) \, exp \left( - \eta \vert t \vert  \right).
\end{equation}
Besides the original Hamiltonian in (\ref{ecs1}), the complete Hamiltonian should include   $V_{ext}$ and the part of the disorder potential 
($\Delta  V(\bs r)=  W^{\dag} \,  V (\bs r) W$)    that was previously subtracted (see \Eq{potren}).  Hence 
the  total  Hamiltonian $H_T$ is written as 
\begin{equation}\label{hamilcomp}
 H_T = H + V_{ext}  + \Delta  V(\bs r)  \, .
\end{equation}
The disorder  potential  $ \Delta  V(\bs r ) $  will  induce  broadening effects, and it  will be lately included. 
Then, the time-evolution for the    density matrix  $\rho(t)$   obeys 
the von Neumann equation 
\begin{equation} \label{vn1}
i \hbar \frac{\partial \rho  }{\partial t}  =  \yav{H + V_{ext}  , \rho} \, . 
\end{equation}
Within the linear regime     $\rho$  is split  in the sum   $\rho = \rho_0 + \Delta \rho$. The  zero order 
term $ \rho_0$ must satisfy 
 \begin{equation} \label{vn0}
 i \hbar \frac{\partial \rho_0  }{\partial t}  =  \yav{H , \rho_0} ,
 \end{equation}
  the conditions required to solve this equation will be established below. The first order deviation $ \Delta \rho$
then obeys 
 \begin{equation} \label{vn2}
 i \hbar \frac{\partial \Delta \rho  }{\partial t}  =  \yav{H  , \Delta  \rho}  +   \yav{ V_{ext}  , \rho_0} \, . 
 \end{equation}
We shall now apply to this equation the three transformation that were  utilized in the previous section in order 
to solve the  \Sh equation, hence   in agreement with \Eq{3trans},  $\tilde {\Delta \rho}$ is defined as 
\begin{equation} \label{dmtran}
  \tilde {\Delta \rho}  (t)   = U^\dag_I(t-t_0)   \exp\{i H_0 t\}  W(t)  \, {\Delta \rho(t)} \,  W^\dag (t)  \exp\{-i H_0 t\}   U_I(t-t_0) \, . 
\end{equation}
In terms of the transformed  density matrix $  \tilde {\Delta \rho}  (t) $,  \Eq{vn2} becomes
 \begin{equation}\label{vn3} 
 i \hbar \frac{\partial   \tilde {\Delta \rho}  }{\partial t}  =  \yav{\tilde{V}_{ext} , \tilde{ \rho}_0 },  
 \end{equation}
where $\tilde{V}_{ext}$ and $ \tilde{ \rho}_0$ are the    external potential and quasi-equilibrium density matrix 
transformed  in the same manner as  $\tilde {\Delta \rho}$ is transformed in  \Eq{dmtran}. The transformed  
quasi-equilibrium density matrix is assumed to have the form 
$ \tilde{ \rho}_0 = \sum_\mu \vert \mu \rangle f({\cal E} _\mu) \langle \mu \vert ,$
where $f({\cal E} _\mu)$  is the usual Fermi function and ${\cal E} _\mu$ the Landau-Floquet levels (see appendix \ref{apenA}). 
 It is straightforward to verify that  this selection guarantee that the 
quasi-equilibrium condition in  (\ref{vn0}) is verified.  The justification for selecting a  Fermi-Dirac distribution in the quasi-energy states is presented in the appendix (\ref{apenC}).   It is shown,  that under experimental  conditions 
($\tau_\omega \ll \tau_{tr} \ll \tau_{in} $ ), the  elastic and inelastic relaxation processes can be neglected as compared to the external field effects. The solution of the  Boltzmann equation  yields,  for a weak microwave intensity,   a Fermi-Dirac distribution in the quasi-energy states.  The expectation value of \Eq{vn3} in the $\vert \mu \rangle$  base can now  be  easily calculated using Eqs. (\ref{3trans}),  (\ref{potext}) and (\ref{dmtran}), solving   the resulting equation with the initial  condition $ {\Delta \rho}  (t)  \to 0$ as $t \to -\infty$, yields for $t < 0$ 
\begin{eqnarray}\label{emdm}
  {\brakete{\mu}{ { \tilde {\Delta \rho}  (t) } }{\nu}}     &=&
  {\brakete{\Psi_\mu}{ {  {\Delta \rho}  (t) } }{\Psi_\nu}}      \nonumber \\
&=&  \frac{e {\bs E}_0}{2} \cdot  \int_{-\infty}^t \left[    
  \frac{e^{i(\Omega - i \eta) t^\prime}}{\Omega}  f_{\mu \nu}  {\brakete{\Psi_\mu}{ {\bs \Pi} (t^\prime)}{\Psi_\nu}}  + \left(\Omega \to - \Omega \right) \right],
\end{eqnarray}
where  the definition  $f_{\mu \nu} =  f({\cal E} _\mu) - f({\cal E} _\nu)$ was used.
Substituting the expectation value for the momentum operator given in    (\ref{emopm}), the integral in the the previous equation 
is easily  performed. The current density to first order in the external electric field  can now be calculated from 
  $\langle  {\bs J} (t, {\bs r} ) \rangle  = Tr \left[   \tilde {\Delta \rho}  (t)   \tilde {\bs J} (t)  \right] $, the resulting expression represents the local      density current. Here we are concerned with the macroscopic  conductivity tensor that relates the spatially and time averaged  current density  
  $ {\bs j } = \left( \tau_\omega V \right)^{-1} \int_0^{\tau_\omega} dt \int d^2x \langle  {\bs J} (t, {\bs r} ) \rangle$ to the averaged electric field; here $\tau_\omega = 2 \pi / \omega$.
 The macroscopic conductivity  can now  be worked out,  results for  the dark and microwave-induced  conductivities  are quoted:
\begin{equation}\label{condd}
 {\bs \sigma}^D_{xi}  =   \frac{e^2 \omega_c^2}{i 4 \hbar} \sum_{\mu \nu} \left\{    \frac{ f_{\mu\nu}}{\Omega}
  \left[ \frac {a_i \mu \delta_{\mu , \nu +1} } {{\cal E} _{\mu\nu} +\Omega - i \eta } 
  + \frac {b_i \nu \delta_{\mu , \nu -1} } {{\cal E} _{\mu\nu} +\Omega - i \eta }  \right]        + \left(\Omega \to - \Omega \right)   \right\},
 \end{equation}
\begin{equation}\label{condw}
 {\bs \sigma}^\omega_{xi}    =    \frac{e^2 \omega_c^2}{i 4 \hbar} \sum_{\mu \nu} \left\{    \frac{ f_{\mu\nu}}{\Omega}
  \sum_l \frac{\Delta_{\mu\nu}^{(l)} (i) \Delta_{\nu\mu}^{(-l) }(x) }{{\cal E} _{\mu\nu} + \omega l + \Omega - i \eta}
   + \left(\Omega \to - \Omega \right) \right\}.
 \end{equation}
 In these expressions   the external electric field points along the $x$-axis.
 Hence, setting $i =x$ or $i=y$  the 
 longitudinal and Hall conductivities  can be selected. The denominators on the R.H.S. of the previous equations 
 can be related to the advanced and  retarded  Green's functions   $G^{\pm}_\mu ({\cal E} ) = 1/\left({\cal E}  -{\cal E} _\mu  \pm i \eta  \right) $. To make further progress the real and absorptive parts of the 
 Green's functions are separated  taking the limit  $\eta \to 0$  and using 
 $lim_{\eta \to 0} \, 1/({\cal E}  -  i \eta) = P 1/{\cal E}   + i \pi \delta({\cal E} )$, where $P$ indicates the principal-value integral. As
 usual  the real and imaginary  parts contribute to the Hall and  longitudinal conductivities respectively.
  In what follows  details of the calculations  are presented  for the   longitudinal microwave-induced conductivity, the corresponding dark conductivity expressions as well as the Hall microwave-induced conductance  are quoted  in the appendix (\ref{apenD}). Implementing the previous considerations  and  inserting a $\delta$ function, the longitudinal microwave-induced conductivity takes the form  
   \begin{eqnarray}\label{condw2}
 {\bs \sigma}^\omega_{xx}    =   -\frac{e^2 \omega_c^2}{4 \hbar} \sum_{\mu \nu} \sum_l  \int d{\cal E}  \delta( {\cal E}  - {\cal E} _\mu)  
  \vert  \Delta_{\mu\nu}^{(l)} (x) \vert^2  \bigg\{    \frac{ f({\cal E}  + \omega l + \Omega )  -  f({\cal E} ) }{\Omega}
 &&  Im \, G_\nu ({\cal E}  + \omega l + \Omega )    \nonumber \\
 &&  +    \left(\Omega \to - \Omega \right)  \bigg\}  , 
 \end{eqnarray}
where $  Im \, G_\nu ({\cal E} )  = \frac{1}{2 i} \left[  G^+_\nu ({\cal E} ) -   G^-_\nu ({\cal E} ) \right]$. 
The static  limit with respect to the external field is obtained taking 
   $\Omega \to 0$.   In the case of the impurity assisted contribution an additional average over the 
 impurity distribution has to be carried out, it is  assumed that the impurities are no correlated,  utilizing the explicit expressions for the velocity matrix elements in (\ref{defdelta}),  the final result 
 for the averaged microwave induced longitudinal conductance is worked out as 
 \begin{eqnarray}\label{condLw}
\langle  {\bs \sigma}_{xx} ^\omega \rangle  &=&   \frac{e^2 }{\pi \hbar}  
\int d{\cal E}     \sum_{\mu \nu}  \sum_l   Im G_\mu \left({\cal E}   \right)B^{(l)}   \left({\cal E}  ,{\cal E} _\nu \right) \bigg\{    
   \omega_c  \vert \rho_1\vert^2  
\delta_{\mu\nu}  \left( \delta_{l,1}  +\delta_{l,-1} \right)                  \nonumber \\
&+ &     n_{imp} \, l^2_B \,  \int \frac{d^2 q}{(2 \pi)^2}    \, \, 
 q_y^2 \, \, \bigg{\vert} {\tilde J}_l\left( \vert \Delta \vert \right)  V( \bs q)  D_{\mu\nu} (\tilde{q}) \bigg{\vert}^2\nonumber \bigg\} ,\\
 \end{eqnarray}
 where $  n_{imp}$ is the  two dimensional impurity density  and the following  function has been defined   
 \begin{equation} \label{derspec} 
 B^{(l)} ( {\cal E} ,  {\cal E} _\nu) =  -\left[  \frac{d}{d{\cal E} _0}  \left\{ \left[ f( {\cal E} + l \omega + {\cal E} _0)-  f ( {\cal E} )\right]  
 Im \, G_\nu ( {\cal E}  + l \omega    + {\cal E} _0) \right\}
  \right]_{{\cal E} _0= 0}. 
 \end{equation}
The photoconductivity  in (\ref{condLw}) has  a first contribution that depends on the $\rho_1$ 
     factor (independent of the  impurity concentration),  it represents the direct cyclotron resonance heating, arising   
     when the $W(t)$ transformation is applied to the momentum operator (see \Eq{tranv}). The impurity induced contribution 
     (second row in  (\ref{condLw})) takes into account the dynamics produced by the magnetic and microwave 
     fields, combined with the resonant effect of  the impurities;  the information is contained in the complete wave function in 
     \Eq{3trans}. 
     
  The  previous expression  would present a singular  behavior that is  an artifact of the $\eta \to 0$ limit. This problem is solved by including the  disorder broadening effects. A simple phenomenological prescription is dictated by simply retaining a finite value of 
   $\eta$ that is related  to the  quasiparticles lifetime ($\eta = 2\pi/ \tau_s$)  \cite{mahan,sinova}. According to this prescription 
   the density of states (DOS) of the $\mu$ level would have a Lorentzian  form $Im \, G_\mu({\cal E} ) = 
    \frac{\eta / 2 \pi}{ \left( {\cal E}  - {\cal E} _\mu \right)^2 + \eta^2/4 }$. A more formal procedure requires to  calculate 
    the broadening produced by,  the so far neglected, part of the disorder potential $  \Delta  V(\bs r)$ (see \Eq{hamilcomp}); fortunately as explained in the appendix  (\ref{apenE}) the calculation becomes equivalent to that carried out  by Ando \cite{ando1} and 
    Gerhardts \cite{gerh}, so the density of states for the $\mu$-Landau level   can be represented by  a Gaussian-type form \cite{ando2}
 \begin{equation} \label{denst}
 Im \, G_\mu({\cal E} ) = \sqrt{\frac{\pi}{2 \Gamma_\mu^2}}  \exp{\left[ - ({\cal E}  - {\cal E} _\mu)^2 /(2 \Gamma_\mu^2) \right] } , 
 \hskip1.5cm    \Gamma^2_\mu = \frac{ 2\beta_\mu  \hbar^2 \omega_c  }{ (\pi \tau_{tr})} .
 \end{equation}   
The parameter  $\beta_\mu $ in  the level width  takes into account the difference  of the transport scattering time determining the mobility $\mu$,  from the single-particle  lifetime. In the case of short-range scatterers  $\tau_{tr} = \tau_s$ and  $\beta_\mu =1 $. An expression for $\beta_\mu$,  suitable for numerical evaluation,   that applies for the long-range screened potential in (\ref{coefp2})  is given in appendix (\ref{apenE}).  $\beta_\mu$  decreases for higher Landau levels.   This property becomes essential  to generate NRS, because they only  appear   for   a   narrow $\Gamma_\mu$,  a condition that is satisfied around the Fermi  level in the case of   large filling factors.
      
     Eqs. (\ref{condLw},\ref{derspec}) contain the main ingredients that explain the huge increase observed in  the longitudinal  conductance (and resistance),   when the material is irradiated by microwaves.  In the standard expression for the Kubo formula there are no Floquet  replica contribution, hence $\omega$ can be set to zero in (\ref{derspec}), if that is   the case $B^{(l)}$ becomes proportional  to the energy derivative  of the Fermi distribution, that  in the $T \to 0$ limit  becomes of the form $\delta ({\cal E}  - {\cal E} _F)$, and the conductivity is positive definite  depending  only on  those states lying  at the Fermi level.  On the other hand,  as a result of the periodic structure induced by the  microwave radiation,  $B^{(l)}$  contains a second  contribution proportional to
   the derivative of the density of states:   $   \frac{d}{d{\cal E} }  Im \, G_\nu ( {\cal E}  + l \omega )$.   Due to the oscillatory structure of the density of states, this extra contribution takes both  positive and negative values. According to \Eq{denst} this second term (as compared to the first one) is proportional to  the electron mobility, hence  for sufficiently  high mobility the new contribution dominates leading to  negative resistance states. The former observation becomes fundamental, because in agreement with experiment,  our calculations show that NRS can only be induced in very high mobility  samples (see \Fig{figure3}). 
   
  As it was mentioned in section  (\ref{model})  the present   method applies in general if   ${\tilde V} ({\bs r})$  
can be decomposed in its Fourier expansion (\ref{potimp}) $e.g.$:
finite wave-length corrections, lattice periodic potential,  impurity scattering, etc. 
The microwave radiation by itself  only produces transitions between adjacent Landau levels (first term on the R.H.S of \Eq{condLw})
leading to the cyclotron peak.  In the case of  a  periodic potential the resulting spectrum will be dominated  by the region 
$ q \approx 2 \pi / a$, where $a$ is the lattice parameter; for the experimental conditions $\lambda_B \gg a$ and the
 contribution is  negligible.  So we are led to analyze the impurity assisted mechanism as a plausible scenario 
 to explain the strong oscillatory structure of the magnetoresistance.

 \section{Results.}\label{results}
 
 The single-particle and transport relaxation rates induced by disorder are given by \cite{davies}
 \begin{equation}\label{relrat}
\rral {\begin{array}{ll}\    \frac{1}{\tau_s}  \\
  \frac{1}{\tau_{tr}}    \\ 
\end{array}  }    
  =  \, n_{imp}  \, \frac{m^*}{ \pi \hbar ^3 k_F} \,  \int_0^{2 k_F} d  q  \,\,  \,\,  \frac{\vert V (q)  \vert^2  }{ \sqrt{1 - (q/2k_F)^2}} \,   \times \, 
\llal{\begin{array}{ll}\  1  \\
\frac{q^2}{2 k_F^2}  \\
\end{array}} \, ,
\end{equation}
where $V(q)$ is the Fourier transform of the impurity potential (\ref{potimp}). Remarkably, we have a consistent formalism in which: 
$(i)$ the  photoconductivity  (\ref{condLw}), $(ii)$ the  relaxation rates (\ref{relrat}), and $(iii)$  the level broadening 
Eqs. (\ref{denst}) and (\ref{broad2}) can all  be consistently calculated once 
   $ V( q)$  has been  specified.
   
 For  neutral impurities the potential can be represented by a short range  delta interaction,  the coefficient in \Eq{potimp} corresponds to a constant   that can be selected as  $ V( q)  =  2 \pi \hbar^2  \alpha / m^*.$
    The expression in \Eq{relrat}  is readily calculated  to   yield the same value for the single-particle and transport relaxation rates
     \begin{equation}\label{nltimes}
    \frac{1}{\tau_{tr}^{(N)}} \, = \,   \frac{1}{\tau_{s}^{(N)}}  \, =  \, \frac{4 \pi^2 \hbar    }{m^*} \, \alpha^2 \,  n^{(N)}_{imp}  \, ,
      \end{equation} 
the upper index $N$ labels the  neutral impurity case. 
     The evaluation of the photoconductivity  (\ref{condLw}) requires in general a  time consuming numerical integration. However, for moderate values of the microwave  radiation the   transitions are dominated by  single  photon exchange, in the neutral impurity case   a very precise analytical approximation can be  explicitly  worked out; see  appendix (\ref{apenF}).
    
 For charged impurities the Coulomb potential is   long-range   modified by the screening effects. Although electron motion is restricted to 2-dimensions, the electric field is three dimensional and there are  contributions from the impurities localized within 
    the doped layer of thickness $d$. The screened potential can then be represented in momentum  coordinates by the expression \cite{davies} 
  \begin{equation}  \label{coefp2}
 V( \vec  q)  = \frac{\pi\hbar^2 }{m^* } \frac{e^{-q d} }{ 1 + \frac{q}{q_{TF}}},  \hskip2.5cm   q_{TF} = \frac{e^2 m^*}{2 \pi \epsilon_0 \epsilon_b \hbar^2}  ,   
    \end{equation}  
where the Thomas-Fermi (TF) approximation is implemented in order to calculate  the dielectric function. Here $\epsilon_b$ represents the relative permittivity of the surrounding media. 
The expression in (\ref{coefp2}) corresponds to a screened potential, that in real space has  a  $r^{-3}$ decay for large $r$.
  The rates  in \Eq{relrat} can be evaluated numerically, however an accurate analytical results is obtained observing that 
  the decaying exponential in (\ref{coefp2}) causes the integral to die off for $ q >> 1/ \vert d \vert$, and the upper limit in 
  the integral can therefore be set to infinity. Additionally for the relevant parameters (see below)  the following conditions are observed  $k_F >> 1/  \vert d \vert $ and $q_{TF} >> 1/  \vert d  \vert $; consequently,  it is reasonable   to drop the factor $q/q_{TF}$ in the denominator of  (\ref{coefp2}) and replace the square root in the 
  denominator of \Eq{relrat} by unity.    These simplifications yields  for the transport relaxation rate
   \begin{equation}\label{cltimes}
    \, \frac{1}{\tau_{tr}^{(C)}}   \, =  \,  \frac{ \pi \hbar }{  8 m^* \left(k_F d  \right)^3  }   \,  n^{(C)}_{imp} \, ,
      \end{equation} 
      the upper index $C$ labels the  charged  impurity case.
 As expected,  for charged remote impurities the single-particle lifetime  differs from the transport lifetime, the approximated relation reads  $\tau^{(C)}_{tr} \approx (2 k_F d )^2 \tau^{(C)}_s$.

\begin{figure} [hbt]
\begin{center}
\includegraphics[width=4.5in]{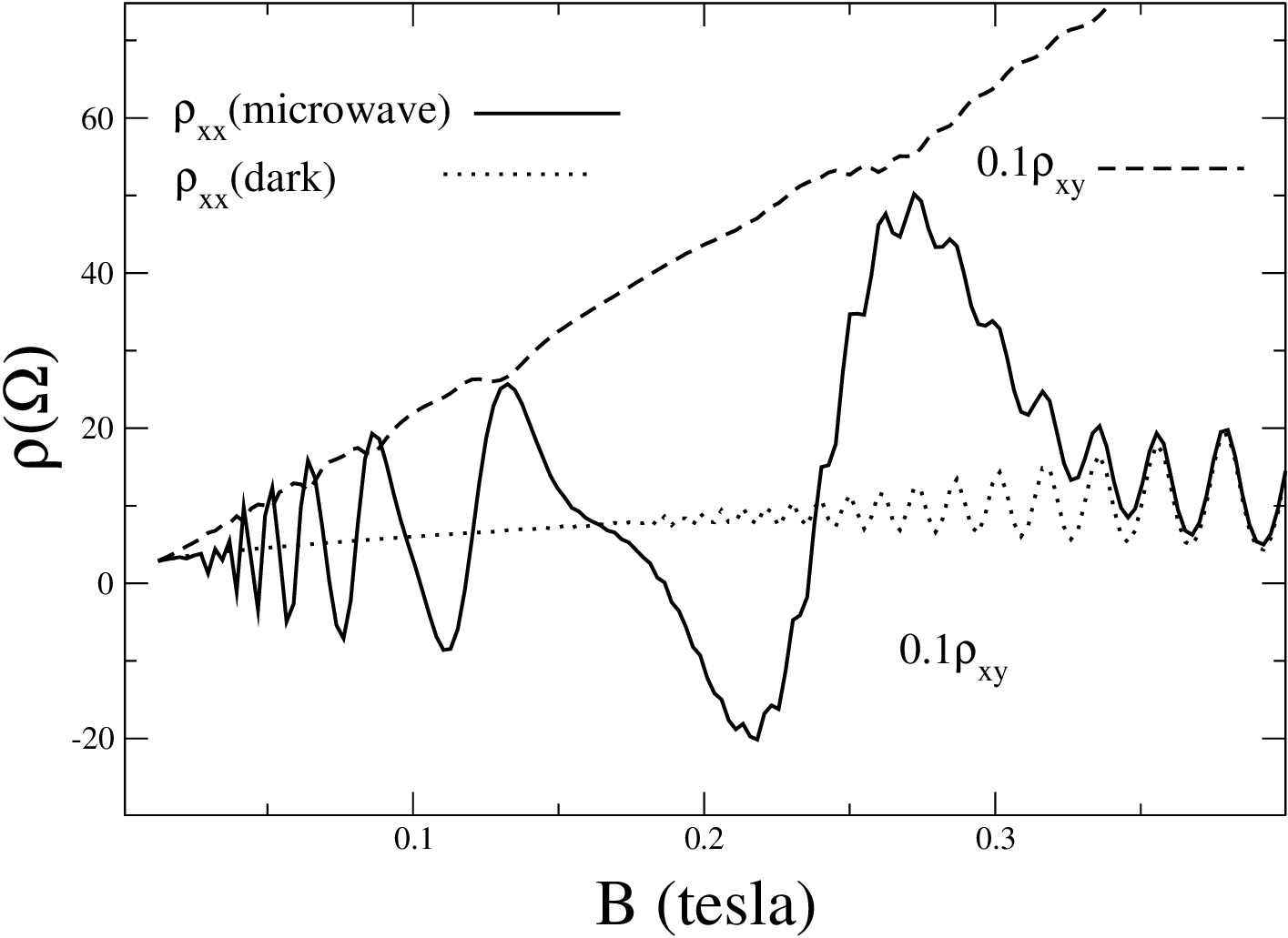}
\end{center}
\caption{ Longitudinal resistance, both  total (continuos line) and dark (dotted line) as a function of the magnetic field.  The figure also 
includes  the Hall resistance (dashed line, $\rho_{xy}$  is rescaled by a factor $1/10$).  Results corresponds to neutral impurity scattering  obtained with the approximated solution (appendix \ref{apenF}) and the selected parameters are: $ \mu \approx 0.25 \times 10^7 cm^2/V  s$,  $T \approx 1\,  K$,  $f  = 100 \, Ghz$,  $ \vert \vec E \vert  \approx 2.5 \, V/cm$,  $\alpha^2 \, n^{(N)}_{imp}  =   5 \times 10^{6} \, cm^{-2}$.
The values of the other  parameters used in the calculations  are discussed   in the text. }
\label{figure1}
\end{figure}

The parameter values have been selected    corresponding to reported experiments   \cite{mani1,zudov2}  in ultraclean  $GaAs/Al_xGa_{1-x} As$ samples:  effective  electron mass $m^* = 0.067 \, m_e$,     relative permittivity   $\epsilon_b \approx 13.18 $, fermi energy ${\cal E} _F = 10 \, meV$, electron  mobility $ \mu \approx 0.1-2.5 \times 10^7 cm^2/V  s$, electron density $n = 3 \times 10^{11} cm^{-2}$,
microwave frequencies $f = 50-100 \, Ghz$,  magnetic fields in the range $0.05 - 0.4 \, \, Tesla$ and temperatures $T \approx 0.5-2.5 \,  K$. The  reported specimen is an   $5 \, mm \times 5 \, mm$ square.  Typical  microwave power is $10-40 \, mW$, however it is estimated \cite{zudov1} that the  microwave power that impinges on the sample surface is of the order  $ 100-200 \, \mu W$,  hence the  microwave  electric field intensity is estimated as  $ \vert \vec E \vert  \approx 1-3 \, V/cm$.   Using these values, it is  verified that  the weak-overlapping condition holds: $\omega_c \, \tau_{tr}  \sim 100-1000$.

\begin{figure} [hbt]
\begin{center}
\includegraphics[width=4.5in]{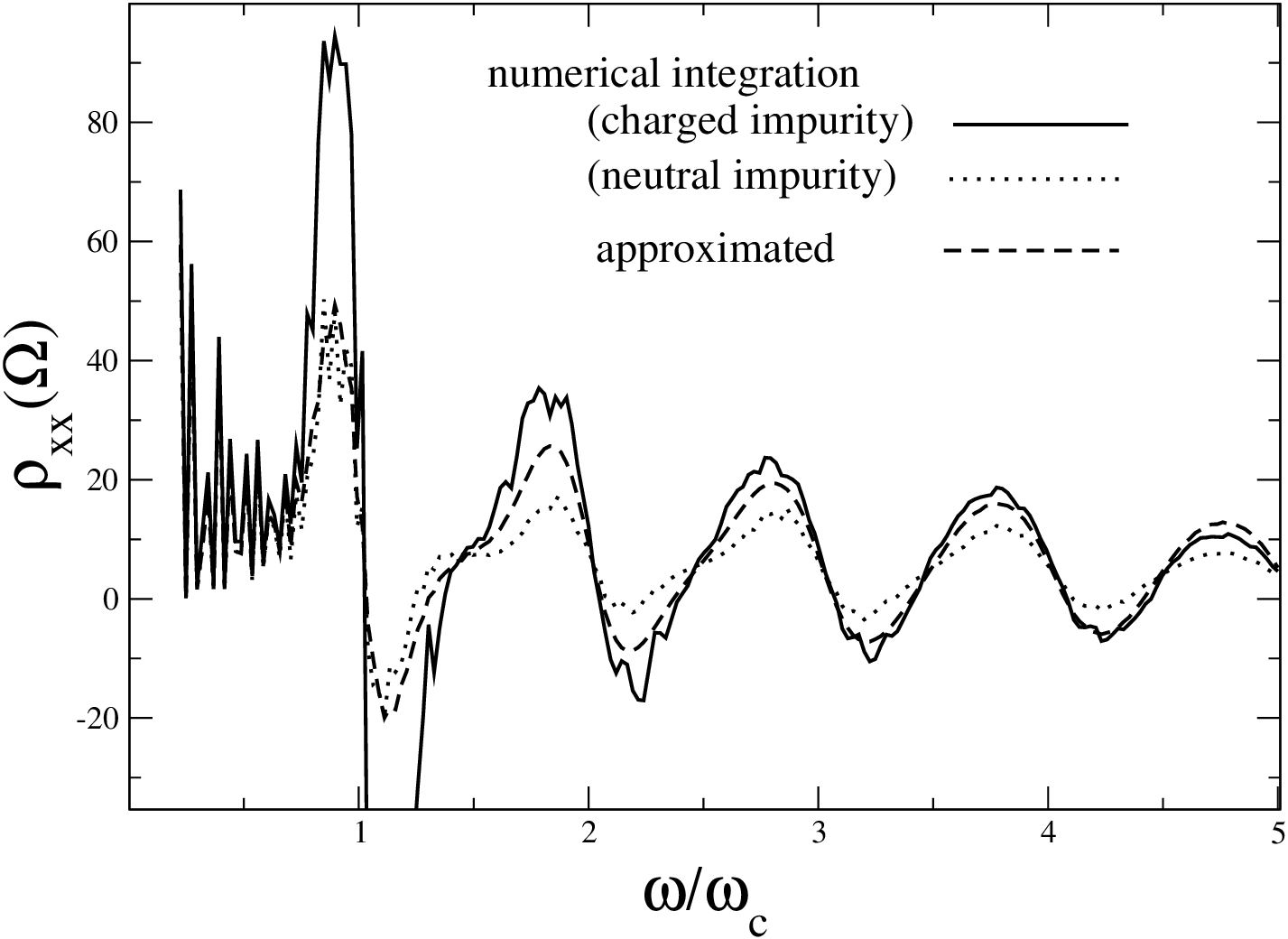}
\end{center}
\caption{ Longitudinal resistance as a function of $\epsilon= \omega/\omega_c$ for neutral and charged impurities. Results for neutral 
impurities   are obtained from the numerical integration (dotted line) and also using  the analytical approximation  discussed in  appendix {\ref{apenF})  (dashed line) with the parameters: 
$ \mu \approx 0.25 \times 10^7 cm^2/V  s$,  $\alpha^2 \, n^{(N)}_{imp}  =   5 \times 10^{6} \, cm^{-2}$. 
The continuos line corresponds to the charged impurity case with parameters $ \mu \approx 2.5 \times 10^7 cm^2/V  s$,  $ \, n^{(C)}_{imp}  =  1.5 \times 10^{11} \, cm^{-2}$. 
The other parameters are the same as in figure \Fig{figure1}.
} }
\label{figure2}
\end{figure}

\begin{figure} [hbt]
\begin{center}
\includegraphics[width=4.5in]{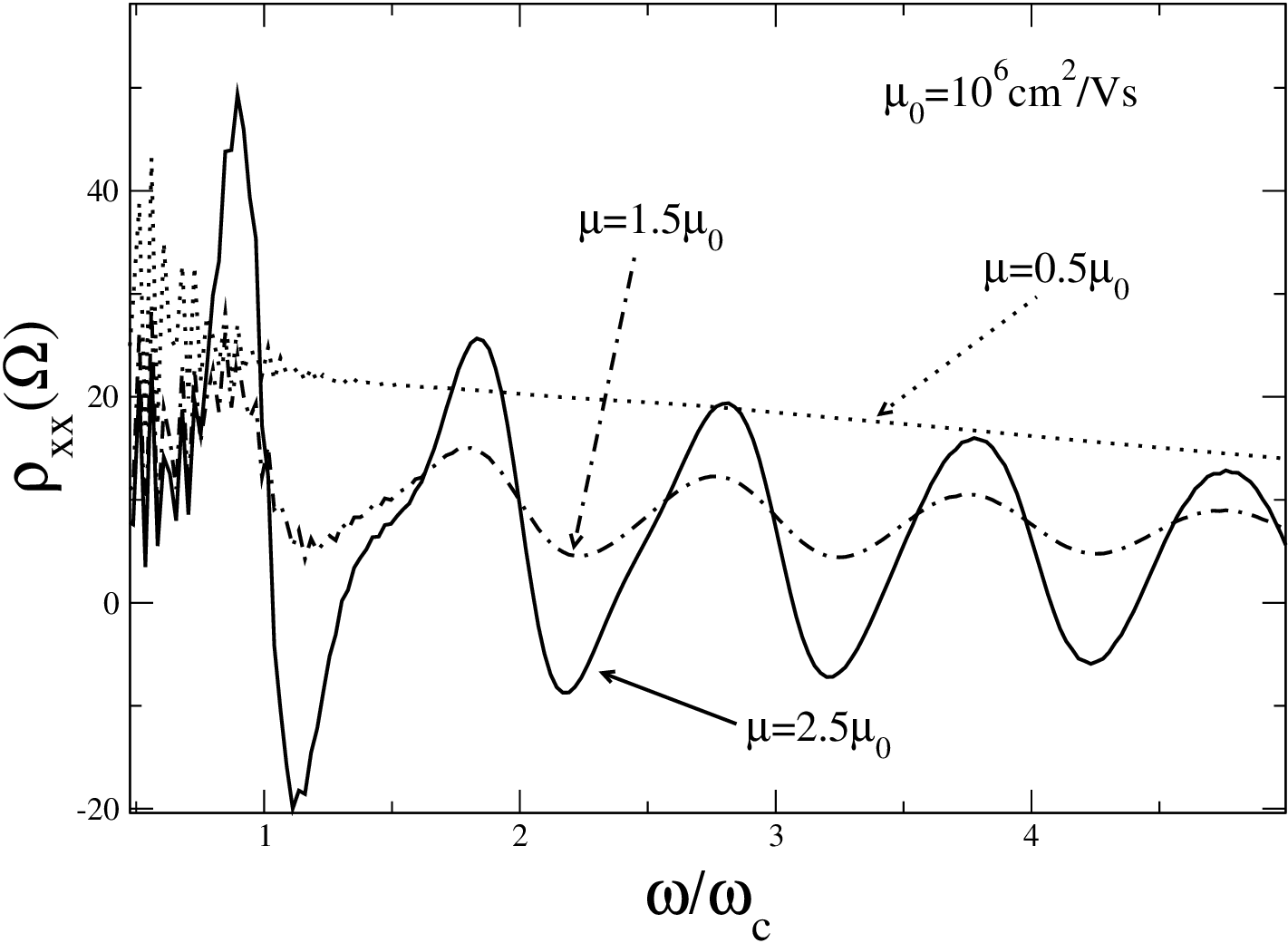}
\end{center}
\caption{  Longitudinal resistivity   as function of  $ \epsilon = \omega / \omega_c $ for neutral impurity scattering and three  values of the electron mobility:  $\mu \approx 0.5  \times 10^6 cm^2/V  s$ (dotted line), $\mu \approx 1.5  \times 10^6 cm^2/V  s$ (dash-dotted line),   and $\mu \approx 2.5  \times 10^6 cm^2/V  s$ (continuos line).
   In  the two  former cases the oscillations follow a pattern with minima at $\epsilon =j + \delta $, and maxima at  
    $\epsilon =j -  delta  $, adjusted with $\delta \approx 1/5$. NRS only appear when  $\mu >  \mu_{th}  \sim 1.5  \times 10^6 cm^2/V  s$.  The values of the  other parameters are the same as in figure \Fig{figure1} }
\label{figure3}
\end{figure}

\begin{figure} [hbt]
\begin{center}
\includegraphics[width=4.5in]{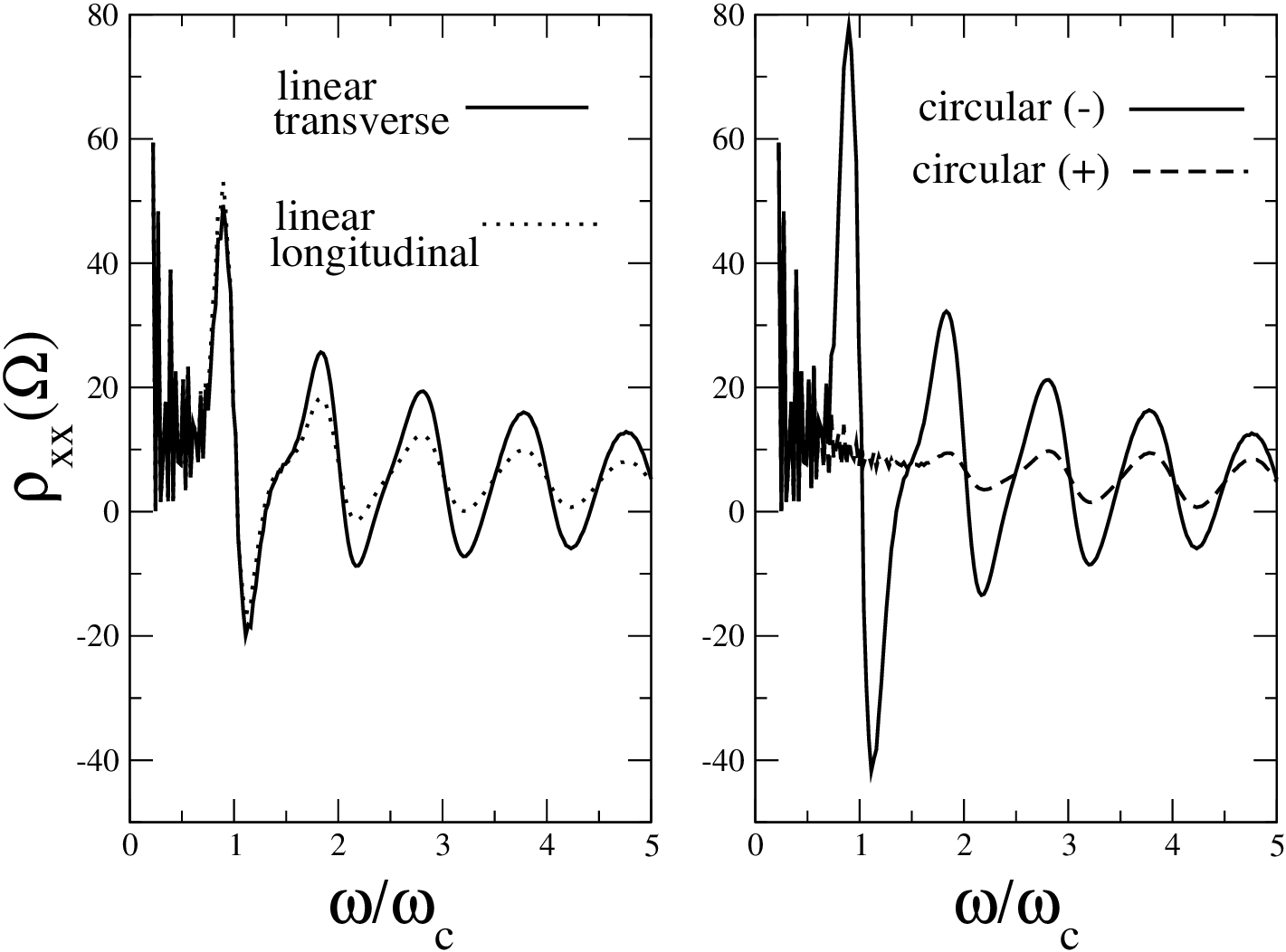}
\end{center}
\caption{ Longitudinal resistance $\rho_{xx}$ for neutral impurity scattering and  various   microwave $E-$field polarization's with respect to the current.  In figure $(a)$ the continuos and dotted  lines correspond to  linear   transverse and longitudinal  polarization's   respectively. 
Figure $(b)$  shows results for circular polarization's:  left-hand (continuos line) and  right-hand  (dashed line). The values of the   parameters are the same as in figure \Fig{figure1} }
\label{figure4}
\end{figure}

\begin{figure} [hbt]
\begin{center}
\includegraphics[width=4.5in]{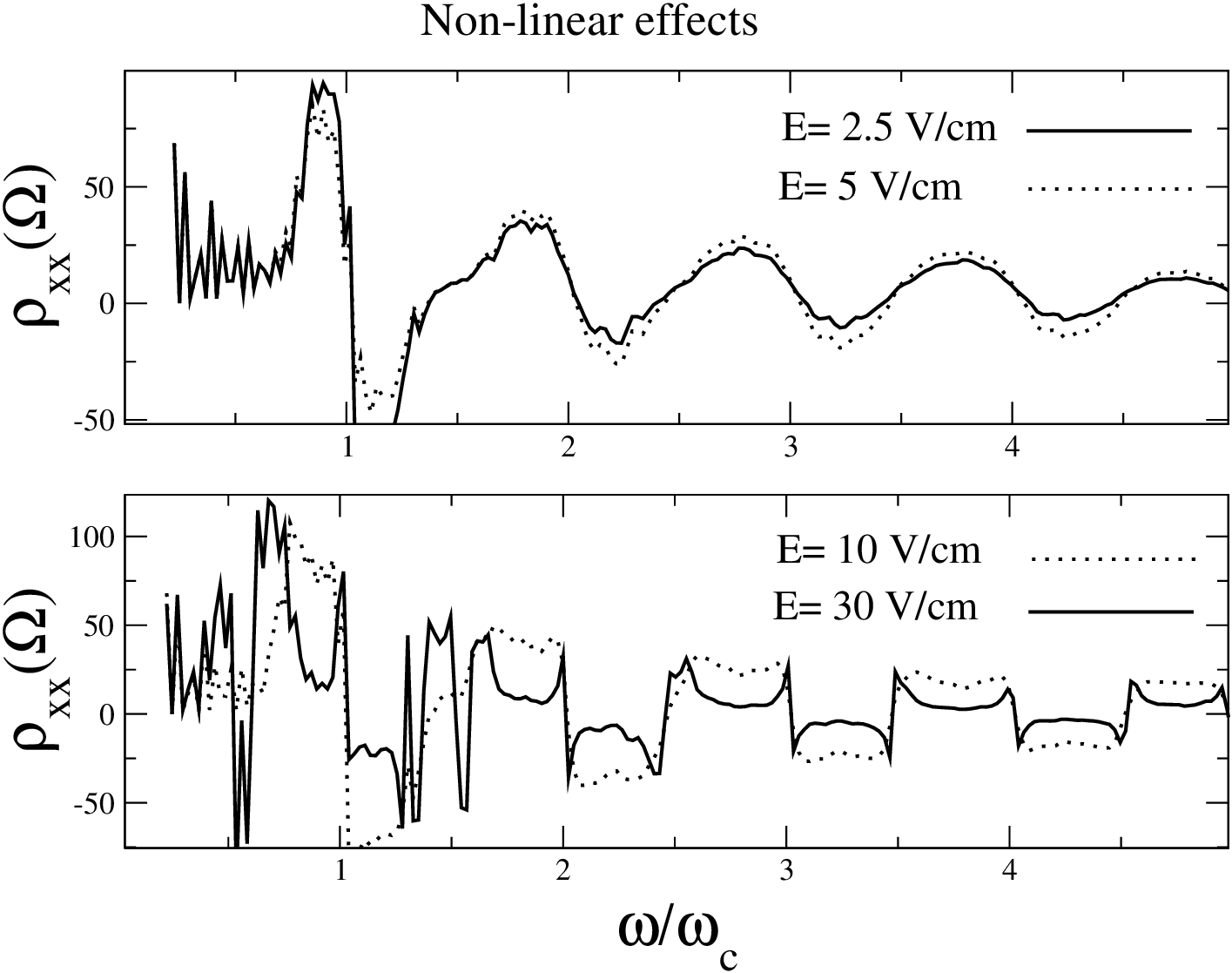}
\end{center}
\caption{ Nonlinear effects in the longitudinal resistance $\rho_{xx}$ for charged impurity scattering. 
Figure (a) includes the $\rho_{xx}$. $v.s.$ $\epsilon=  \omega/\omega_c$ plots for 
electric filed intensities of $\vert \vec E \vert = 2.5 \, V/cm$ (continuos line) and  $\vert \vec E \vert = 5 \, V/cm$ (dotted line). 
Figure (b) displays results for $\vert \vec E \vert = 10 \, V/cm$ (dotted line)   and $\vert \vec E \vert = 30 \, V/cm$ (continuos line). 
The parameter values are: $ \mu \approx  2.5 \times 10^7 cm^2/V  s$,  $T \approx 1\,  K$,  $f  = 100 \, Ghz$,  
 $ \, n^{(C)}_{imp}  =  1.5 \times 10^{11} \, cm^{-2}$.   }
\label{figure5}
\end{figure}

   Recalling that   $\mu = e \tau_{tr}  /m^*$, one can use    Eqs. (\ref{nltimes}) and  (\ref{cltimes})  to determine the values of $n_{imp}$ corresponding  to neutral or   charged  scatterers respectively. For example assuming   $ \mu \approx 2.5 \times 10^6 cm^2/V  s$, one estimates  for neutral scattering  $\alpha^2 \, n^{(N)}_{imp}  \approx   1 \times 10^{7} \, cm^{-2}$. 
   Although $\alpha$ and $ n^{(N)}_{imp}$ are not separately fixed,  one notices that the 
   condition for weak disorder potential as compared to the Landau energy can be be expressed as $ V(q)/(l_B^2 \hbar \omega_c) = 2\pi \alpha \ll 1$; $e.g$  if $\alpha \sim 0.01$, then $n^{(N)}_{imp} \sim   10^{11} \, cm^{-2}$.
   For charged impurities and taking a value for   the separation  $d$  between the impurity and the 2DES  as   $ d \approx 20 \, nm$, yields 
   $n_{imp}^{(C)} \approx  1.5  \times 10^{11} \, cm^{-2}$.   In this case the weak disorder condition
   takes the form $ V(q)/(l_B^2 \hbar \omega_c) \sim \pi \exp ( - 2 \pi d / l_B) \ll 1$, that is  satisfied. 
   A final remark is related  to  the radiative electron decay $\Gamma_{rad}$ that determines the direct electron response to the microwave 
   excitation (see \Eqs{soleqm} and (\ref{defdelta2})), following  reference \cite{mikha}   $\Gamma_{rad}$  is related to 
   the radiative decay  width   that is interpreted as coherent dipole re-radiation of  electromagnetic waves by the oscillating 2D
   electrons excited by microwaves. Hence, it is given by  $\Gamma_{rad} = n e^2 /\left( 6 \epsilon_0  m^*  c \right)$,  using the values of $n$ and $m^*$ given above it yields  $ \Gamma_{rad} \approx 0.38 \, meV $.

   Adding the dark and microwave  induced  conductivities, the total longitudinal $\sigma_{xx} = \sigma_{xx}^D +
   \langle \sigma_{xx}^w \rangle$, and Hall $\sigma_{xy} = \sigma_{xy}^D +
   \langle \sigma_{xy}^w \rangle$ conductivities are obtained. It should be pointed out that the interference between the dark and microwave contributions exactly cancels. The corresponding resistivities are obtained from the expression 
   $ \rho_{xx} =\sigma_{xx}/\left(  \sigma_{xx}^2 + \sigma_{xy}^2 \right)$ and  $ \rho_{xy} =\sigma_{xy}/ \left( \sigma_{xx}^2 + \sigma_{xy}^2 \right)$. The relation $\sigma_{xy} \gg \sigma_{xx}$ holds in general,  hence it follows that $\rho_{xx} \propto \sigma_{xx}$,  and the longitudinal resistivity  follows the same oscillation  pattern as  that of  $\sigma_{xx}$. The plots of the total longitudinal and Hall resistivities as a function of the magnetic field intensity are displayed in \Fig{figure1}. Whereas the Hall resistance presents the expected 
   monotonous behavior, the longitudinal resistance shows a strong oscillatory behavior 
   with distinctive NRS. The behavior of the complete $\rho_{xx}$ is contrasted with the dark contributions that presents only the expected Shubnikov-de-Hass oscillations. 

 \Fig{figure2}  shows a  comparison  of the  longitudinal resistivity  as function of   $ \omega / \omega_c  $ obtained for  the case of neutral impurity scattering  using both the approximated expression in the appendix (\ref{apenF})  as well as the result of the numerical integration 
   \Eq{condLw};  the electron  mobility is selected as  $ \mu = 0.25 \times 10^7 cm^2/V  s$. The approximated analytical result  shows a good agreement with the one obtained from the numerical integration. It should  be remarked that the approximated expression includes only one-photon exchange processes, while the numerical result
   includes the possibility of multi-photon exchange, hence it is concluded that for the selected electric field intensity 
   ($ \vert \vec  E \vert \sim 2.5 \, V/cm$), the  one-photon processes   dominate. 
   Results are also presented for the case of charged impurity scattering but for  $ \mu  = 2.5 \times 10^7 cm^2/V  s$. In spite 
   of the very different nature of the two physical processes and that the mobility is increased by an order of magnitude in the charged case, it is observed that the results for the  neutral and charged cases are very similar.  The   similarity of both results is based on: $(i)$    the increase in the mobility is compensated by the factor $\beta_\mu$ in (\ref{denst}) giving a similar broadening value, $(ii)$ although for the  neutral   case $V(q)$ is constant over  all the $q$-range of integration, whereas  for charged case $V(q)$ varies   according to the expression in (\ref{coefp2}),  the  integral in (\ref{condLw}) is  dominated  by the region in which  $q \approx 2 \pi / l_B$.

  One of the puzzling  properties of the observed huge  magnetoresistance oscillations is
     related to the fact that they appear only in samples with an electron mobility exceeding  a   threshold value $ \mu_{th}  $.  The  phenomenon is absent  in samples in  which $\mu$ is   slightly reduced. This behavior is well reproduced   by the present formalism. \Fig{figure3}  displays the $\rho_{xx} \, \, vs. \,\, \epsilon = \omega/ \omega_c $ plot for neutral impurity scattering and   three  selected values of $\mu$.  For $\mu \approx 0.5  \times 10^6  cm^2/V  s$ the  previously known,  almost linear behavior $\rho_{xx} \propto B$ is clearly depicted.  As the electron mobility  increases to $\mu \approx 1.5  \times 10^6 cm^2/V  s$, the magnetoresistance oscillations are clearly observed; however, NRS only  appear when   the mobility is increased to $\mu \approx 2.5  \times 10^6 cm^2/V  s$. It is observed  that $\rho_{xx}$ vanishes at $\epsilon =j$ for $j$ integer.  The  period and  phase of the oscillations follow a pattern very similar to the one observed in experiments \cite{mani1,mani2},  with minima at $\epsilon =j + \delta $, and maxima at  
      $\epsilon =j -  \delta $, adjusted with $\delta  \approx 1/5$.  It should be pointed out that this  value of $\delta$  depends on the correct representation of the density of states, using a Lorentzian form instead of the Gaussian in (\ref{denst}) would give  $\delta \sim 1/10$ \cite{torkun}.   Similar behavior  is observed  for the charged impurity scattering  case, but with  the mobility threshold increased  approximately by  and order of magnitude $ \mu_{th}  \approx 2.5  \times 10^7 cm^2/V  s$. The precise determination of  $ \mu_{th}  $ depends of course on the selected values of the other parameters, mainly  on  the frequency and microwave intensity. 
    
   The first reported experiments \cite{zudov1,mani1,zudov2,mani2} were carried out for a microwave radiation with transverse polarization with respect  to the longitudinal current flow direction. However, it is clear that the presence of the magnetic field  induces a chirality in the system, the model can  be used  to test these effects.    \Fig{figure4}   shows the results  for different 
   ${\bs E}-$field polarization's with respect to the current. In   \Fig{figure4}a  it is observed that the  amplitudes of  the resistivity   oscillation  are  slightly  bigger for transverse  polarization as compared to longitudinal polarization.  This result is in agreement with the recent experiment \cite{mani2}, in which it is reported that the selection of longitudinal or transverse polarization produces small differences. However, we propose that the more significant signatures will be only observable for circular polarization.  Selecting  negative circular polarization (see \Fig{figure4}b), the oscillation amplitudes get the maximum possible value.  Instead,  for positive circular polarization  an important reduction of the  amplitude  is observed  leading to the  total disappearance  of   the NRS. These results are understood  recalling  that   for negative circular polarization and $ \omega \approx \omega_c$ the  electric field rotates in phase with respect to the  electron cyclotron rotation.  Based on the present results, it will be highly recommended to carry out experiment for circular polarization configurations.

   The present formalism can also be used in order to explore the non-linear regime in which multi-photon exchange  plays an essential role. As the microwave radiation  intensity is increased,  the analytical approximation breaks down and the numerical expression in (\ref{condLw}) with higher  multipole  ($l$) terms needs  to be evaluated. In the explored regime 
   convergent results are obtained  including terms up to the $l=3$ multipole. \Fig{figure5}a  displays  $\rho_{xx}$ $vs.$ 
   $\epsilon $ plots   for  electric field intensities $\vert \vec E \vert = 2.5 \, V/cm $ and $\vert \vec E \vert = 5 \, V/cm $ respectively, the increase on the  field intensity  produces a corresponding increase in the minima and maxima of $\rho_{xx}$, but apart of this, the 
   qualitative behavior in both cases is similar.  A further increase of the electric field intensity to  $\vert \vec  E \vert = 10 \, V/cm $ and
    $\vert \vec E \vert = 30 \, V/cm $ (\Fig{figure5}b), 
   takes us to the nonlinear regime in which  a qualitatively new behavior is observed. For $\epsilon > 2$  the same 
 NRS  are observed, however the width of these regions increase to include practically all the range from 
   $\epsilon = j$ to  $\epsilon = j + \frac{1}{2}$. Notice that the negative resistance minima has not a monotonous dependence on $\vert \bs{E} \vert$, in fact for the strongest field intensity the minima approaches zero.  
     Remarkably,  for $\vert \vec  E \vert = 30 \, V/cm $ and  $\epsilon < 2$, new negative resistance states associated with transitions by two 
     microwave photons are  observed near     $\epsilon =  \frac{1}{2}$ and  $\epsilon =  \frac{3}{2}$.   The minima of these states 
     are centered at  $\epsilon_{min} =0.52 $  and  $\epsilon_{min} =1.52 $    respectively. 
     Evidence of ZRS  associated with multi-photon processes have been already observed by Zudov $et \, al.$ 
     (\cite{zudov1}); they reported structures with maxima near $\epsilon = 1/2$ and $\epsilon = 3/2$, and the corresponding minima centered around  $\epsilon_{min}= 0.67$ and $\epsilon_{min} = 1.68$ respectively. Dorozhkin \cite{doro}    and Willett  $et \, al.$  (\cite{willett}) have also reported $\rho_{xx}$ minimum associated with $\epsilon = 1/2$. 
     Although the exact position of the minima and maxima of $\rho_{xx}$    observed in \Fig{figure5}b are not localized at the same position 
     reported by Zudov $et \, al.$   (\cite{zudov1}), the general pattern is very similar, supporting the interpretation as multi-photon processes.
     Clearly, a more systematic analysis and further experimental studies are necessary.

    Comparison with some other theoretical work is obliged.  Previous work  in references 
    \cite{ry1,ry2,durst,vavilov} analyzed  the effects of  the microwave radiation  on the electron scattering by impurities in  the presence of a magnetic field. Durst $et \, al$  \cite{durst} consider an out of equilibrium calculation, instead here 
   a quasi-adiabatic approximation is implemented,   assuming that the   system is thermalized in those states characterized by the Landau-Floquet spectrum.  The similarity between  some results in the  present work and those of Durst $et \, al$  \cite{durst},  suggest that 
   departure from equilibrium is not significant for the studied  phenomenon.  The present formalism  extends and explores  the impurity assisted photoconductivity mechanism  in  detail.    In this  model  the same  disorder potential determines the broadening of the Landau levels, as well as the wave function that is used  to evaluate the velocity matrix element. These matrix element  are incorporated into a Kubo-like formula that takes into account the Floquet structure of the system.  As previously mentioned, there are alternative models in which the leading  contribution arises from the modification  of the electron distribution function induced by the microwave radiation.
    According to Dorozhkin \cite{doro} the negative resistance phenomena has its origin in a local population inversion that produces a change of sign of the $\left( \partial f / \partial {\cal E}  \right) $ term that appears in the conductivity.  Although possible, the inversion of population   requires rather strong microwave powers, which were not achieved in the experiments \cite{mikha}. Indeed, the inversion population is expected to be produced when the microwave energy  exceeds the Fermi energy  $   (eE) ^2 /(m^* \omega^2)  > E_F$ (see appendix \ref{apenC}),  clearly   the estimated value for the threshold electric field $E_{th} \sim 1000 \, V/cm$ highly exceeds the experimental microwave  fields $  E \sim1-5 V/cm$. An interesting alternative explanation based on the modifications that the microwave  radiation produces in the distribution function was  recently presented by  Dmitriev $et \, al.$  \cite{dmi1} and Kennett $et \, al.$$et \, al.$  \cite{kennett}. In these publications it is assumed that the inelastic-scattering processes  give  the dominant contribution to  the collision term of the  kinetic equation.  However, as explained in the appendix (\ref{apenC})  under  experimental conditions    $\tau_\omega \ll \tau_{tr} \ll  \tau_{in} $, and certainly the inelastic processes can be safely ignored  as compared to the elastic processes.  In fact we have presented an argument for a first approximation in which the distribution function is determined only by the  microwave effects.  It  may be interesting for a future work, to add to the present formalism the effects  that  elastic processes produces to the  distribution  function.  In any case,  we consider that the present results taken together with those of references   \cite{ry1,ry2,durst,vavilov}  consolidate the explanation of the photoconductivity oscillations and negative resistance states in terms of the 
    microwave-disorder mechanism.

 \section{Conclusions.}\label{conclu}
   
We have considered  a model to describe the photoconductivity of a 2DEG  subjected to a magnetic 
field.  We presented a thoroughly discussion of the  method that allow us to take into account  the Landau 
and microwave contributions in a non-perturbative exact way, while  the impurity scattering effects are treated perturbatively. The method exploits  the   symmetries  of the problem:  the exact  solution of the  Landau-microwave dynamics (\ref{tran1}) is obtained in terms of the electric-magnetic generators (\ref{trancan})  as well the solutions  to the  classical  equations of motion (\ref{lagrang}).  The  spectrum  and Floquet modes are explicitly worked out.   In our model,  the Landau-Floquet  states act coherently with respect to the  oscillating field of  the impurities, that   in turn induces  transitions between these levels.  Based on this formalism, a  Kubo-like   formula is provided,  it takes into account the oscillatory Floquet structure of the problem. 
It should be stressed that the disorder potential is conveniently split (see Eqs. (\ref{potren}),  (\ref{hamilcomp}) ) in such a way that it contributes both to the matrix  elements of the velocity operator, as well as to the broadening of the Landau levels. Hence,  we have a consistent formalism in which: 
$(i)$ the  photoconductivity  (\ref{condLw}), $(ii)$ the  relaxation rates (\ref{relrat}), and $(iii)$  the level broadening 
Eqs. (\ref{denst}) and (\ref{broad2}) can all  be consistently calculated once 
 the disorder potential  has been  specified.

The   expression for the  longitudinal photoconductivity \Eq{condLw}  contains the main ingredients that explain the huge increase observed in the experiments.  As explained in section (\ref{kubo}),    the standard expression for the Kubo formula  at low temperature is dominated  by   the states   near to the Fermi level.  On the other hand,  as a result of the periodic structure induced by the  microwave radiation the term   $B^{(l)}$  contains a second  contribution proportional to the derivative of the density of states:
   $   \frac{d}{d{\cal E} }  Im \, G_\nu ( {\cal E}  + l \omega )$.   Due to the oscillatory structure of the density of states this extra contribution takes both  positive and negative values. According to \Eq{denst} this second term is proportional to the electron mobility, hence  for sufficiently  high mobility the new contribution dominates leading to  negative resistance states. 
This allows us to explain  one of the puzzling  properties of the observed huge  magnetoresistance oscillations, related to the fact that they appear only in samples with an electron mobility exceeding  a threshold.  This result is well 
 reproduced by the present model, for the selected parameters,    NRS  emerge when the condition 
 $\mu \ge  2.5  \times 10^6 cm^2/V  s$ (short-range disorder), and  
$\mu \ge  2.5  \times 10^7 cm^2/V  s$  (long-range disorder) are satisfied.   The oscillations follow a pattern with minima at $ \epsilon  =j + \delta $, and maxima at    $ \epsilon  = j -  \delta $, adjusted with $\delta  \approx 1/5$. These results are in reasonable good 
 agreement with the observation of Mani $et \, al.$  \cite{mani1,mani2}, they  reported a similar pattern with $\delta \approx 1/4$.

    An interesting prediction of the present model is related to  polarization effects 
  that could be possible observed in future experiments. While the results for the cases of linear transverse or longitudinal polarization's
  show small differences, the selection of circular polarized radiation leads to significant signatures. 
    The maximum possible value  for  the oscillation amplitudes of $\rho_{xx}$ appears for negative circular polarization.  Instead,   positive circular polarization   yields an important reduction  on  the oscillation   amplitudes   and  the  total disappearance  of   the NRS.  This result can be understood, if one recalls    that for  negative circular polarization and $ \omega \approx \omega_c$ the  electric field rotates in phase with respect to the  electron cyclotron rotation.  The present results call for the importance  to carry out experiment with  circular polarization configurations.

An analysis  was   presented in order   to explore the non-linear regime in which multi-photon exchange  plays an essential role. The results suggest the existence of new  NRS  (which 
are expected to develop into ZRS) near $\epsilon= 1/2$ and $\epsilon= 3/2$, these states  correspond to two-photon exchange processes  and are in reasonable agreement  with the reported experimental results.

 Some final remarks are related to the limitations and possible extensions of the present work.  In a first approximation we have not included the contribution of the elastic processes to the kinetic equation that determines the electron  distribution. However,  it will be interesting to extend the present calculations to include, not only the dynamical effects produced by the impurity  on the electron wave function, but also the  modifications that they produce in the distribution function.


 \section{Appendices.}\label{apen}
 
 \subsection{Microwave driven Landau problem.}\label{apenA}
 
  \Eq{trancan} defines  a canonical transformation from the variables $\{ t, x, y ; p_0,p_x,p_y \}$ to $\{ Q_0, Q_1, Q_2; P_0,P_1,P_2 \}$, 
  in terms of the new variables the   \Sh equation (\ref{ecs1})  (without impurity potential) takes the form 
  \begin{equation}\label{ecsa}
  P_0 \Psi = \left[ \hbar \omega_c \frac{Q_1^2 + P_1^2}{2} + e l_B E_x \left(Q_1 - P_2 \right) + e l_B E_y \left(Q_2 - P_1 \right)  \right] \Psi   \, .
 \end{equation}
 The action of the transformation $W$ defined in \Eq{opw} over the $(Q_\mu,P_\mu)$  variables can be easily calculated as:
 \begin{eqnarray}\label{tranope}
WQ_0 W^\dag &=& Q_0 ,  \nonumber \\
  WP_0 W^\dag &=& P_0 + \dot \eta_1 Q_1 + \dot \zeta_1 P_1 +\dot \eta_2 Q_2 + \dot \zeta_2 P_2
 - \dot \zeta_1 \eta_1 - \dot \zeta_2 \eta_2 + { \cal L}  ,                  \nonumber \\
 WQ_1 W^\dag &=& Q_1 + \zeta_1 , \hskip2.0cm        WP_1 W^\dag = P_1 -  \eta_1 ,                    \nonumber \\
 WQ_2 W^\dag &=& Q_2 + \zeta_2 , \hskip2.0cm  WP_2 W^\dag = P_2 - \eta_2     \, . 
\end{eqnarray}
It can be verified that when the $W$   transformation is applied to \Eq{ecsa},  the second and third terms in the right hand side  exactly cancel with all the terms  that appear in the expression for  $WP_0 W^\dag $ (except $P_0$)  if  ${\cal L} $ is identified with the Lagrangian in \Eq{lagrang} an   the 
functions $\eta_i$ and $\zeta_i$ satisfy the  differential equations that follows from the variation of  ${\cal L} $: 
 \begin{eqnarray}\label{ecmov}
\dot \eta_1 - \omega_c \zeta_1 &=&e l_B E_x , \hskip2.5cm    \dot \zeta_1 + \omega_c \eta_1 = - e l_bE_y    ,                  \nonumber \\
 \dot \eta_2 &=& e l_B E_y ,  \hskip2.5cm  \dot \zeta_2 = - e l_B E_x                \, . 
\end{eqnarray}
 But,  these are exactly the   classical equations of motion that follow  when  the variational principle  is applied to the Lagrangian in \Eq{lagrang}. Hence the  $W $ operator    transforms  the  \Sh equation  (\ref{ecsa})   to the Landau eigenvalue  problem with the Hamiltonian given  in \Eq{tran1}.

 For the electric  field  consider the expression in \Eq{vecpot}.
 It is then straightforward  to obtain the solutions to  \Eqs{ecmov}, adding   a damping term that takes into account the radiative decay of the quasiparticle, they read 
 \begin{eqnarray}\label{soleqm}
 \eta_1 &=& e l_B  E_0  \, Re \left[\frac{-i \omega \epsilon_x + \omega_c \epsilon_y  }{\omega^2 - \omega_c^2 + i \omega \Gamma_{rad} }
     e^{i \omega t} \right] ,   \hskip2.0cm     \eta_2 = e l_B E_0 \,  Re \left[\frac{\epsilon_y e^{i\omega t}}{i \omega} \right] ,                  \nonumber \\
\zeta_1 &=& e l_B  E_0 \, Re \left[\frac{ \omega_c \epsilon_x + i \omega \epsilon_y  }{\omega^2 - \omega_c^2 + i \omega \Gamma_{rad}  }
     e^{i \omega t} \right] ,  \hskip2.0cm      \zeta_2 = -e l_B E_0 \, Re \left[\frac{\epsilon_x e^{i\omega t}}{i \omega} \right] .
\end{eqnarray}
According to the Floquet theorem the wave function  can be written as
 $\Psi (t) = \exp \left(- i {\cal E} _\mu  t \right) \phi_\mu(t)$, where $ \phi_\mu(t)$ is periodic in time, 
 $i.e.$ $ \phi_\mu(t + \tau_\omega ) =  \phi_\mu(t)$.  From   \Eq{opw} it is noticed that  the transformed wave function  $\Psi^W = W \Psi$ contains the phase factor $\exp \left( i \int^t   {\mathcal L}  d t^\prime \right) $. It then follows  that the quasienergies and the  Floquet modes can be deduced if we add and subtract to this  exponential  a term of the form $\frac{t}{\tau} \int_0^\tau {\mathcal L} dt^\prime$. Hence, the quasienergies can be  readily read off
\begin{equation}\label{enerflo}
 {\cal E} _\mu ={\cal E} _\mu^{(0)} +   {\cal E}_{rad} \, ;  \hskip0.8cm  
   {\cal E} _\mu^{(0)} = \hbar \omega_c \left( \frac{1}{2} + \mu  \right) ,   \hskip0.8cm  {\cal E}_{rad} = 
   \frac{e^2 E_0^2 \left[ 1 + 2 \omega_c {Re(\epsilon_x^* \epsilon_y) }/ \omega\right] } {2 m^* \left[ \left(\omega - \omega_c \right)^2 + \Gamma_{rad}^2 \right]}, 
\end{equation}
here  ${\cal E} _\mu^{(0)}$ are   the usual Landau energies, and the induced  Floquet energy shift is given by the microwave energy 
$ {\cal E}_{rad} $.   
The corresponding time-periodic Floquet modes in the 
  $(P_1,P_2)$ representation are given by 
\begin{equation}\label{wf2}
\Psi_{\mu,k } (P) = \exp\{- i \sin\left(2 \omega t \right) F(\omega) \} \phi_\mu(P_1) \delta(P_2 - k) \, , 
\end{equation}
here  $\phi_{\mu}\fues{P_1}$ is the harmonic oscillator function  in the 
$P_1$ representation
\begin{equation}\label{hofun}
\phi_{\mu}\fues{P_1}=\brak{P_1}{\mu}=
\frac{1}{\sqrt{\pi^{1/2}2^{\mu}\mu !}}
e^{-P_1^2/2}H_{\mu}\fues{P_1} \, , 
\end{equation}
and  $H_{\mu}\fues{P_1}$  is the Hermite polynomial and the function $F(\omega)$ is given as 
\begin{equation}
F(\omega) = \frac{\omega_c}{\omega}\left(\frac{e E_0 l_B }{\omega^2 - \omega_c^2 } \right)^2 \left[  \omega^2 - \omega_c^2  + 2 \omega^2 \epsilon_x^2
- 2 \omega_c^2  \epsilon_y ^2 + \frac{Re(\epsilon_x^* \epsilon_y) }{\omega \omega_c}  \left(2 \omega^4 - \omega^2  \omega_c^2 
+ \omega_c^4 \right)  \right]. 
\end{equation}
The wave function (\ref{wf2}) depends on the Landau ($\mu$), and center guide ($k$) indexes; however the spectrum (\ref{enerflo})  is degenerate wit respect to $k$.  It is important to notice that   $F(\omega)$  appear in  the wave function phase, that depends only on  time,  hence  its contribution to the expectation value of the momentum  operator   cancels  exactly. Thus,  contrary to what it is  claimed  in reference \cite{shi}, the effect of the Floquet dynamics (without including an extra effect such as impurity scattering) can not account for the  explanation of the ZRS  observed  in recent experiments.

 \subsection{Impurity induced transitions.}\label{apenB}

In this appendix we consider the first order solution of the evolution operator $U(t)$ given 
by 
\begin{equation}\label{evolU}
U(t)  = 1 - i \int_{-\infty}^{t}  dt^\prime \left[ W^{\dagger}(t^\prime)  {\tilde V} (\bs r) W(t^\prime)   \right]_I \, .
\end{equation}
The effect of the transformation induced by the $W$  operator over the impurity potential can be easily evaluated
considering the effect over the Fourier decomposition of $V({\bs r})$ given in \Eq{potimp}.  Recalling that the $x$ and $y$ 
coordinates  are written in terms of the new variables $(Q_1,P_1,Q_2,P_2)$ by means of \Eq{invtran}, and 
utilizing the transformation  properties  of the $(Q_i,P_i)$ operators in (\ref{tranope}), it is readily  obtained 
\begin{eqnarray}
& &   W^{\dagger}(t)  \exp\{i {\bs q \cdot \bs r } \} W(t)   =  
   \exp\{  i l_B \left(q_x P_2 - q_y Q_2  \right)  \}    \times \nonumber \\
    & &   \exp\{-   i l_B \left( q_x Q_1 - q_y P_1  \right)  \}  \times
     \exp\{  i l_B \left[q_x \left(\zeta_1 + \eta_2 \right) +  q_y \left(\zeta_2 + \eta_1 \right) \right]  \}  .
\end{eqnarray}
Using  \Eqs{soleqm}  the third exponential in the previous equation can be recast in a compact form 
as $\exp \{ - i Re \left( \Delta \exp ( i \omega t )  \right) \} $, with $\Delta$ given in \Eq{defdelta2}; this expression 
  can be expanded  as  \cite{abra}
 \begin{equation}
\exp \{ - i Re \left( \Delta \exp ( i \omega t )  \right) \} = \sum_{-l = \infty}^{l = \infty}
 \left( \frac{\Delta}{i \vert \Delta \vert } e^{i \omega t } \right)^l \, J_l\left(\vert \Delta \vert  \right),
\end{equation}
 with $J_l$ the  Legendre polynomials. For the the second exponential  notice that once that 
 $Q_1$ and $P_1$  are replaced by the raising and lowering  operators given in \Eq{opasc}, one is lead to 
 evaluate the matrix elements of the operator $D\fues{ \tilde{q}} = \exp \left(  \tilde{q} a_1^\dag  -  \tilde{q}^*  a_1 \right)$
 that generates coherent Landau states. A calculation yields 
\begin{equation}\label{laguerres}
D^{\nu \mu}\fues{ \tilde{q}}=\braket{\nu}{ D\fues{ \tilde{q}}}{\mu}
=e^{-\frac{1}{2}\abs{  \tilde{q}}^2}
\llal{\begin{array}{ll}\fues{- \tilde{q}^{*}}^{\mu-\nu}
\sqrt{\frac{\nu!}{\mu!}}L^{\mu-\nu}_{\nu}
\fues{\abs{ \tilde{q}}^2}, \,\,\,\,& \mu >\nu, \\
 \tilde{q}^{\nu-\mu}\sqrt{\frac{\mu!}{\nu!}}
L^{\nu-\mu}_{\mu}\fues{\abs{ \tilde{q}}^2},
& \mu <\nu,\\
\end{array}}
\end{equation}
where $L^{\mu}_{\mu}$ are the generalized Laguerre polynomial.   With all these provisions the matrix element of the solution of the evolution operator in (\ref{evolU}) 
  can be worked out  as 
  \begin{equation}
 {\brakete{ \mu}{U(t)  }{ \nu}} = \delta_{\mu\nu} - \sum_l \frac{e^{i \left( {\cal E} _{\mu\nu} + \omega l \right)t}}{{\cal E} _{\mu\nu} + \omega l  - i \eta } \, C^{(l)}_{\mu\nu},
\end{equation}
the explicit expression for $ C^{(l)}_{\mu\nu}$ was given in (\ref{defc}).

 \subsection{Microwave-driven distribution function.}\label{apenC}
Within  the time relaxation approximation the Boltzmann equation  can be written as 
   \begin{equation}\label{boltz}
\frac{\partial f }{\partial t } + \frac{\partial f }{\partial \bs{p} } \cdot \left( e \bs{E} + e \bs{v} \times \bs{B}  \right) = - \frac{f - f_F }{\tau_{tr}} 
- \frac{f - f_F }{\tau_{in}}  ,
\nonumber 
 \end{equation}
 where  $f_F$  is  the Fermi-Dirac distribution and  we distinguish between the elastic  rate $\tau_{tr}^{-1}$ and inelastic or energy relaxation rate $\tau_{in}^{-1}$. 
 Under  experimental conditions:  $\tau_\omega \ll \tau_{tr} \ll  \tau_{in} $, and certainly the inelastic processes can be safely ignored. Furthermore,  due  to the $ac$-electric field (\ref{vecpot}),   the L.H. S. of the previous equation is estimated to be of order $f / \tau_\omega$; hence, in a first approximation  the elastic scattering contribution can also be neglected. The resulting Vlasov equation has the exact solution $f(\bs{p}, t) = f_F \left( \bs{p}  - m^* \bs{v} (t)\right) $, where the velocity 
 $ \bs{v} (t) \equiv \left(\dot \eta_1, \dot \zeta_1 \right) $ solves exactly the same classical equations of motion as given in (\ref{ecmov}), and the initial condition is  selected 
 as $f  \to f_F$  as the external electric field is switched-off.  In particular it is  verified that 
 $ m^* \vert \bs{v} (t) \vert^2/2 =  {\cal E}_{rad} $  coincides with  the Floquet energy shift  produced by the microwave radiation(\ref{enerflo}). 
 The steady-state distribution,  evaluated at  the Landau energy ${\cal E} ={\cal E} _\mu^{(0)}$,  is obtained by averaging $ f_F \left( \bs{p}  - m^* \bs{v} (t)\right) $ over the oscillatory period
   \begin{equation}\label{boltz2}
\langle f_F \rangle = \frac{1}{\tau_\omega} \int_0^{\tau_\omega} f_F \left( {\cal E} _\mu^{(0)}  +  {\cal E}_{rad}+
2 \, \cos \omega_c t   \,  \sqrt{ {\cal E} _\mu^{(0)} \,   {\cal E}_{rad}} \,  \right)  \, dt 
\nonumber . 
 \end{equation}
 For the experimental conditions it is verified that  $  {\cal E}_{rad}  \ll {\cal E} _\mu^{(0)}  $, thus  expanding to first order 
one finds   $\langle f_F \rangle   \approx f_F \left( {\cal E} _\mu^{(0)}  +  {\cal E}_{rad} \right) =   f_F( {\cal E} _\mu   ) $. Hence, it is verified that a  rapid relaxation of the Fermi distribution to the quasi-energy states is a  reasonable assumption. The arguments presented in this appendix have  been introduced  by  Mikhailov \cite{mikha} in order to explore the possibility that the microwave radiation leads to a population inversion;  however,  it is concluded  that it  would require a rather high   microwave intensity 
 $  {\cal E}_{rad} > {\cal E} _F$.

   \subsection{Dark and Hall  conductivities.}\label{apenD}
   In section (\ref{kubo}) it was explained in detail the method  to obtain the final expression for the 
   microwave induced magnetoresistance  \Eq{condLw}. 
   Working along a similar procedure the expression for the  remaining conductivities are worked  from  
    equations (\ref{condd})  and (\ref{condw}).
   First we quote the  longitudinal dark conductance 
      \begin{equation}
 {\bs \sigma}^D_{xx}   =  \frac{ e^2 \omega_c^2}{\pi \hbar}   \sum_\mu \, \mu \int d{\cal E}  \,  Im \, G_\mu\left( {\cal E}   \right)
 \,   \frac{d f }{d {\cal E}  } \,  Im \, G_\mu \left( {\cal E}  + \omega_c \right) \, ,
 \end{equation}
 whereas the dark  Hall conductance  is given by 
   \begin{equation}
 {\bs \sigma}^D_{xy}   =  \frac{ e^2 \omega_c^2 }{\pi \hbar  } \sum_\mu  \, \mu \int d{\cal E}  \, Im \, G_\mu\left( {\cal E}   \right)
 \left[ f\left( {\cal E} _\mu - \omega_c \right) -   f\left( {\cal E}  \right)\right]  \, P\, \frac{1}{\left({\cal E}  - {\cal E} _\mu + \omega_c \right)^2}
 \, , 
 \end{equation}
 where $P$ indicates the principal-value integral. 
   The impurity assisted contributions require  an additional average over the 
 impurity distribution, it is  assumed that the impurities are no correlated. The  final result for the microwave assisted 
 longitudinal conductivity was quoted in \Eq{condLw}. Following a similar procedure the microwave assisted 
 Hal conductivity is calculated to give 
 \begin{eqnarray}\label{condHw}
\langle  {\bs \sigma}_{xy} ^\omega \rangle  &=&   \frac{ e^2 \omega_c^2 }{\pi \hbar}  
\int d{\cal E}     \sum_{\mu \nu}  \sum_l   Im \, G_\mu \left({\cal E}   \right)  
 \left[ f\left( {\cal E} _\nu  \right) -   f\left( {\cal E}_\mu  \right)\right] 
\bigg\{    
\delta_{\mu\nu}  \left( \rho_1 \rho_2^* \delta_{l,1}  + \rho_1^* \rho_2 \delta_{l,-1} \right)                  \nonumber \\
&+ &     n_{imp} \,  \int d^2 q   \, \, 
 T(\bs q) \, \, \bigg{\vert} {\tilde J}_l\left( \vert \Delta \vert \right)  V( \bs q)  D_{\mu\nu} (\tilde{q}) \bigg{\vert}^2\nonumber \bigg\} ,\\
 \end{eqnarray}
 were the function $T(\bs q )$ is defined as
\begin{equation} \label{funaux2} 
T(\bs q ) = \omega_c^3 \,  l_B^2 \, \, \frac{  q_x^2 \, + \,   q_y^2 }
   { \left(  {\cal E}  + \omega l  - {\cal E} _\nu    \right) \,\, {\vert \left({\cal E}  + \omega l -  {\cal E} _\nu    \right)^2 - \omega_c^2 \vert^2}} \, .
 \end{equation}
 
  \subsection{Landau density of states.}\label{apenE}
  
  A  detailed calculation of the density of states incorporating  all the elements that contribute to the 
  system under study is beyond the scope of the present paper.  However, it can be argued that the expression given 
  in \Eq{denst} for the DOS is  expected  to be a reasonable  selection under some consistent  approximations. 
 Le us consider the   Green's function associated with the  Hamiltonian $H + \Delta V$, where $H$ is given  in (\ref{ecs1}) 
 and  $\Delta V =W^\dag V W$ is the subtracted  part of the disorder potential (\ref{hamilcomp}). As explained in section (\ref{kubo}),  the  Kubo formula (\ref{condd},\ref{condw}) was deduced using the wave function 
  obtained after the three transformations in  \Eq{3trans} are applied  to the Landau states. 
Hence, $\Delta V$ is transformed  according to 
$ \tilde {\Delta V}  (t)   = U^\dag_I(t-t_0)   \exp\{i H_0 t\}  W(t)  \, {\Delta V } \,  W^\dag (t)  \exp\{-i H_0 t\}   U_I(t-t_0) $  (see \Eq{dmtran}).
Notice that: $(i)$  the $W$ transformation cancels exactly, $(ii)$ both $\Delta V$ and the first order correction to  $U_I(t-t_0)$ 
are proportional to $V$, hence considering linear terms on $V$ we can set $U_I(t-t_0) \approx 1$, $(iii)$ finally when evaluated  in the 
$\vert \mu \rangle$ base and  neglecting inter-Landau mixing,  the contributions from   $\exp\{-i H_0 t\}$  cancels out. 
Hence $ \tilde {\Delta V} \approx V(\bs{r})$, and the problem under consideration reduces to evaluate the density of states  produced by a magnetic field and a disorder potential  $V(\bs{r})$ of the form given in \Eq{potimp}; but this is precisely the problem considered some ago  time  by  by Ando \cite{ando1} and Gerhardts \cite{gerh}; the density of states is well represented by the Gaussian expression in (\ref{denst}), and the level broadening neglecting couplings between different Landau levels is taken from reference \cite{ando2}: 
 \begin{equation}\label{broad1}
 \Gamma^2_\mu=8*\pi l^2_B  n_{imp} \int \frac{d^2 r }{2\pi l^2_B}  \int \frac{d^2 r^\prime }{2\pi l^2_B }
 V(\bs{r})  V(\bs{r^\prime}) \left[  D^{\mu \mu}\fues{ \vert \bs{r} - \bs{r}^\prime \vert  (\sqrt{2} l_B)} \right]^2,
 \end{equation}
where $D^{\mu \mu}$ is given in (\ref{laguerres}). For the delta  short range scatterers   the previous expression is  readely 
 evaluated yielding the result in (\ref{denst}) with $\beta_\mu=1$. In the case of the charged impurity  disorder, after the substitution 
of  the Fourier decomposition (\ref{potimp}) and using \Eq{coefp2} it is verified that $\Gamma_\mu$ is again given by the expression in \Eq{denst}, but  the factor $\beta_\mu$ is  given by  
\begin{equation}\label{broad2}
 \beta_\mu=16 \pi  \left( k_F \, d \right)^3\int_0^\infty \,  q \, \frac{ \exp{\left[ - \sqrt{8} \frac{d}{l_B}  q\right] }}{\left(1 + \frac{\sqrt{2}\, q}{l_B \,  q_{TF}} \right)^2}
  \left[  D^{\mu \mu}\fues{ q } \right]^2 \,  dq \,. 
 \end{equation}
 Previous analysis of the Landau level broadening were carried out, for example  \cite{ando1},  for a Gaussian potential $ V(\bs{r}) \sim e^{r^2/d^2}$, however as mentioned in section (\ref{results}) the actual situation corresponds to a screened potential that in real space has  a $r^{-3}$ decay for large $r$.  As mentioned in section (\ref{kubo}) the value of $\beta_\mu $ decreases  for higher Landau levels. For example for the selected parameter, we have: $\beta_0 = 108$, $\beta_{30} =14 $, $\beta_{50}  =11 $.

 \subsection{Approximated one-photon exchange photoconductivity.}\label{apenF}
 The microwave induced longitudinal (\ref{condLw}) conductivity requires the 
numerical evaluation  of a time consuming integral given by  
   \begin{equation}\label{inte1}
 S_l =  \int d^2 q     K(\bs q) \bigg{\vert} J_l\left( \vert \Delta \vert \right)  V( \bs q)  D_{\mu\nu} (\tilde{q}) \bigg{\vert}^2\nonumber .
 \end{equation}
However,  if we consider the regime of moderate microwave intensity and assuming neutral impurity scattering,  a 
very useful analytical approximation can be worked out. For neutral impurity scatterers, the potential is assumed
to be of the short range delta form, hence the Fourier coefficient in (\ref{potimp}) is given by  
$ V( \bs q)  =  2 \pi  \hbar^2 \alpha /m^* $. The $D_{\mu\nu} (\tilde{q})$ term contains an exponential 
factor that represents a cut off for large $q$. Then according to \Eq{defdelta2} for moderate values of the microwave 
electric field the $\Delta$ term is small  and the leading contributions arises from the $l = \pm 1$ factors 
that correspond to the single photon exchange contribution. Using the approximation $J_1 (z) \approx z/2$
one is lead to evaluate 
\begin{equation}\label{inte2}
 S_{\pm 1}  =   \frac{2 \pi  \hbar^2  \alpha }{m^*  } \int d^2 q     K(\bs q)   \vert \Delta \vert^2  \vert D_{\mu\nu} (\tilde{q})  {\vert}^2 . 
 \end{equation}
 The angular integration is straightforward, while the integral over the $q = \sqrt{q_x^2 + q_y^2}$ leads, after a  change of 
 variable $\xi = q^2 $ to an integral  of the form 
\begin{equation}\label{inte3}
 \int_0^\infty d \xi  e^{-\xi} \xi^{\mu - \nu + 2 } \left(L_\nu^{\mu - \nu} \right)^2, 
 \end{equation}
that is explicitly evaluated with the help  of the recurrence relation 
$ x L_n^k = (2 n + k + 1)L_n^k - (n + k ) L_{n - 1}^k - (n+1 ) L^k_{n+1}$ and the integral \cite{abra}
 \begin{equation}\label{inte4}
 \int_0^\infty d \xi  e^{-\xi} \xi^{k } \, L_n^k \, L_m^k  =  \frac{(n + k)\, !}{n \, !} \, \delta_{m n}.
 \end{equation}
The final result reads 
  \begin{equation}
  S_{\pm 1}  = 
  \frac{\pi \,  \hbar^2 \,  \alpha \, \omega^2 \, \vert E \vert^2   \,  I_{\mu \nu }    }{8 m^* \omega_c^2 \vert \omega^2 - \omega_c^2 + i \omega \Gamma\vert^2 } 
      \left[ \omega^2 ( 1 + 2 \vert \epsilon_y \vert^2 )   + \omega_c^2 ( 1 + 2 \vert\epsilon_x\vert^2 )  - 8 \omega \omega_c Im\left(\epsilon^*_x \epsilon_y \right)    \right]  ,
  \end{equation}
   where 
  \begin{equation}
 I_{\mu \nu }  =   6 \left( \mu + \frac{1}{2} \right)    \left(\mu + \frac{1}{2}   \right)   +    \left(\mu - \nu  \right)^2 + \frac{1}{2}.
 \end{equation}
These expressions greatly simplifies the numerical calculations, and as it is discussed in section (\ref{results}) 
they  provide  a very accurate  approximation to the exact result.

\acknowledgments We acknowledge the partial financial support endowed by
CONACyT through grants No.   \texttt{42026-F} and   \texttt{43110}.

\newpage

\end{document}